\documentclass[useAMS,usenatbib]{mn2e}
\usepackage{graphicx,fleqn,natbib,amssymb}
\usepackage{epstopdf}
\usepackage{xcolor}
\usepackage{hyperref}
\usepackage{adjustbox}

\DeclareGraphicsExtensions{.eps,.cps,.ps,.eps.gz,.ps.gz,.eps.Z}
\def \deg{^\circ}

\usepackage{dcolumn}
\usepackage{pdflscape}
\usepackage{dblfloatfix}
\usepackage{rotating}
\usepackage{longtable}
\usepackage[flushleft]{threeparttable}
\usepackage{tablefootnote}

\title[{\it Swift}/UVOT GW O3 follow-up]{\it{Swift}/UVOT follow-up of Gravitational Wave Alerts in the O3 era}
\author[Oates et al.]{S. R. Oates$^{1,2,3}$\thanks{E-mail: s.r.oates@bham.ac.uk}, F. E. Marshall$^{4}$, A. A. Breeveld $^{2}$, N. P. M. Kuin$^{2}$, P. J. Brown$^{5}$,
\newauthor  M. De Pasquale$^{6}$, P.A. Evans$^{7}$, A. J. Fenney$^{2}$, C. Gronwall$^{8,9}$, J.A. Kennea$^{8}$, 
\newauthor N.J. Klingler$^{8}$, M. J. Page$^{2}$, M. H. Siegel$^{8}$, A. Tohuvavohu$^{10}$, E. Ambrosi$^{11}$, 
\newauthor S.D. Barthelmy$^{4}$, A.P. Beardmore$^{7}$, M.G. Bernardini$^{12}$, S. Campana$^{12}$, R. Caputo$^{4}$,
\newauthor   S.B. Cenko$^{4,13}$, G. Cusumano$^{11}$, A. D'A{\` i}$^{11}$, P. D'Avanzo$^{12}$, V. D'Elia$^{14,15}$,
\newauthor  P. Giommi$^{15}$, D.H. Hartmann$^{16}$, H.A. Krimm$^{17}$, S. Laha$^{4,18}$, D.B. Malesani$^{19}$, 
\newauthor  A. Melandri$^{12}$, J.A. Nousek$^{8}$, P.T. O'Brien$^{7}$, J.P. Osborne$^{7}$, C. Pagani$^{7}$, K.L. Page$^{7}$,
\newauthor  D.M. Palmer$^{20}$, M. Perri$^{15,14}$, J.L. Racusin$^{4}$, T. Sakamoto$^{21}$, B. Sbarufatti$^{8,12}$, 
\newauthor J.E. Schlieder$^{4}$, G. Tagliaferri$^{12}$\& E. Troja$^{4,22}$ \\
  $^{1}$ School of Physics and Astronomy \& Institute for Gravitational Wave Astronomy, University of Birmingham, B15 2TT, UK\\
  $^{2}$ University College London, Mullard Space Science Laboratory, Holmbury St. Mary, Dorking, RH5 6NT, UK\\
  $^{3}$ Department of Physics, University of Warwick, Coventry, CV4 7AL, UK \\
  $^{4}$ Astrophysics Science Division, NASA Goddard Space Flight Center, Greenbelt, MD 20771, USA\\
  $^{5}$ George P. and Cynthia Woods Mitchell Institute for Fundamental Physics and Astronomy, Mitchell Physics Building, \\Texas A.\&M. University, Department of Physics and Astronomy, College Station, TX 77843, USA\\
  $^{6}$ Department of Astronomy and Space Sciences, Istanbul University, Beyaz{\i}t, 34119, Istanbul, Turkey\\ 
  $^{7}$ School of Physics and Astronomy, University of Leicester, LE1 7RH, UK\\
  $^{8}$ Department of Astronomy and Astrophysics, The Pennsylvania State University, University Park, PA 16802, USA\\
  $^{9}$ Institute for Gravitation and the Cosmos, The Pennsylvania State University, University Park, PA 16802, USA\\
  $^{10}$ Department of Astronomy and Astrophysics, University of Toronto, Toronto, ON, Canada\\
  $^{11}$ INAF -- IASF Palermo, Via Ugo La Malfa 153, I-90146, Palermo, Italy\\
  $^{12}$ INAF -- Osservatorio Astronomico di Brera, Via Bianchi 46, I-23807 Merate, Italy\\
  $^{13}$ Joint Space-Science Institute, Computer and Space Sciences Building, University of Maryland, College Park, MD 20742, USA\\
  $^{14}$ INAF-Osservatorio Astronomico di Roma, Via Frascati 33, I-00040 Monteporzio Catone, Italy\\
  $^{15}$ Space Science Data Center (SSDC) - Agenzia Spaziale Italiana (ASI), I-00133 Roma, Italy\\
  $^{16}$ Department of Physics and Astronomy, Clemson University, Kinard Lab of Physics, Clemson, SC 29634-0978, USA\\
  $^{17}$ National Science Foundation, Alexandria, VA 22314, USA\\
  $^{18}$ Center for Space Science and Technology, University of Maryland Baltimore County, 1000 Hilltop Circle, Baltimore, MD 21250, USA\\
  $^{19}$ DTU Space, National Space Institute, Technical University of Denmark, Elektrovej 327, 2800 Kongens Lyngby, Denmark\\
  $^{20}$ Los Alamos National Laboratory, B244, Los Alamos, NM, 87545, USA\\
  $^{21}$ Department of Physics and Mathematics, Aoyama Gakuin University, Sagamihara, Kanagawa, 252-5258, Japan\\
  $^{22}$ Department of Physics and Astronomy, University of Maryland, College Park, MD 20742, USA\\
}

\begin{document}

\date{Accepted...Received...}
\maketitle
\clearpage
\begin{abstract} 
In this paper, we report on the observational performance of the {\it Swift} Ultra-violet/Optical Telescope (UVOT) in response to the Gravitational Wave alerts announced by the Advanced Laser Interferometer Gravitational Wave Observatory and the Advanced Virgo detector during the O3 period. We provide the observational strategy for follow-up of GW alerts and provide an overview of the processing and analysis of candidate optical/UV sources. For the O3 period, we also provide a statistical overview and report on serendipitous sources discovered by {\it Swift}/UVOT. {\it Swift} followed 18 gravitational-wave candidate alerts, with UVOT observing a total of $424$ deg$^{2}$. We found 27 sources that changed in magnitude at the 3$\sigma$ level compared with archival $u$ or $g$-band catalogued values. {\it Swift}/UVOT also followed up a further 13 sources reported by other facilities during the O3 period. Using catalogue information, we divided these 40 sources into five initial classifications: 11 candidate active galactic nuclei (AGN)/quasars, 3 Cataclysmic Variables (CVs), 9 supernovae, 11 unidentified sources that had archival photometry and 6 uncatalogued sources for which no archival photometry was available. We have no strong evidence to identify any of these transients as counterparts to the GW events. The 17 unclassified sources are likely a mix of AGN and a class of fast-evolving transient, and one source may be a CV.
\end{abstract}


\section{Introduction}
\label{intro}
The detection of the electromagnetic (EM) counterpart to Gravitational Wave (GW) event GW170817 \citep[e.g.,][]{abb17,abb17b,cou17,eva17,gol17} marked our entry into the era of GW-EM multi-messenger astronomy. With observations by a large number of dedicated telescopes and follow-up programs\footnote{To gain an idea of the number and breadth of EM facilities following GW events, we refer the reader to \cite{abb17} which summarises the EM follow-up of GW170817.} such as were implemented at the {\it Neil Gehrels Swift Observatory} \citep[henceforth {\it Swift};][]{geh04}, we expect routine detection of astrophysical sources in both gravitational and electromagnetic waves. A variety of astrophysical phenomena are expected to produce GW signals, including compact binary coalescence (CBC, the coalescence of e.g. binary black hole, BBH; binary neutron stars, BNS; or black hole - neutron star BH-NS), core-collapse supernovae and magnetar flares \citep[e.g][]{abb09}. The detection of GW signals together with their EM counterparts is important as it enables a more complete picture of astrophysical phenomena to be formed. Indeed, a breakthrough occurred when the first BNS GW event was detected, GW170817 \citep{abb17,abb17b}. Associated with this GW event was a weak short Gamma-ray Burst (GRB) 170817A detected by the Fermi and Integral satellites \citep[e.g][]{gol17,sav17} and a bright kilonova \citep[KN; AT2017gfo; e.g][]{and17,arc17,cou17,cho17,cow17,dia17,dro17,eva17,fon17,gal17,hu17,kas17,lipu17,mcc17,nic17,pia17,sha17,sma17,soa17,tan17,uts17,val17,vil17,poz18,vil18}. Days after the GW/GRB event, an X-ray and radio counterpart emerged which suggested the origin was off-axis GRB afterglow emission \citep[e.g][]{hal17,mar17,tro17}. The association of the EM counterpart of GW170817 to the host galaxy NGC 4993 allowed the first application of GWs as standard sirens, measuring the Hubble Parameter using the distance information from the GW signal and the redshift information from the EM signal \citep[e.g][]{abb17b,abb17c,guid17,pal17,can18,lee18,hot19}.

On 1st April 2019, the Advanced Laser Interferometer Gravitational Wave Observatory \citep[LIGO; LIGO Scientific Collaboration][]{aa15} and the Advanced Virgo detector \citep[Virgo; the Virgo Scientific Collaboration;][]{ace15} began the third observing run (“O3”) in search of GW events \citep{GCN:24045}\footnote{On the 25th February 2020, KAGRA commenced science observations, officially joining the international network of GW detectors \citep{kag19}.}. O3 was divided into two segments of 6 months each, separated by a month break: O3a and O3b. O3b was expected to officially end on 30th April 2020 but was cut short a month early due to the COVID-19 pandemic. One notable difference between this run and previous runs was the public release of GW alerts; in O1 and O2, the alerts were only released to the EM follow-up partners. GW triggers detected by the LIGO Scientific Collaboration and Virgo Collaboration (LVC) are assigned several parameters, including a false alarm rate (FAR; characterizing the frequency at which noise with the same strength as the signal is expected to arise), whether the detected signal arose from a CBC or an unmodelled burst, and (for CBC triggers) the estimated distance of the merger and the masses\footnote{The mass estimates of the binary components were not released in the initial announcements.} of the initial compact objects. Automated preliminary notices are announced through Gamma-ray Coordinates Network (GCN) Notices\footnote{https://gcn.gsfc.nasa.gov/} when analysis results in a FAR of less than 1 per 10 months or 1 per 4 years for CBC and unmodelled burst searches, respectively. These notices are quickly followed up with a GCN Circular, released after human vetting, that provides either a confirmation of the GW alert, with an updated sky localization and source classification, or a retraction\footnote{https://emfollow.docs.ligo.org/userguide/index.html}. 

{\it Swift} was designed specifically to detect and follow-up GRBs. However, in the last few years, {\it Swift} has increasingly been used to explore a wide range of transient astrophysical phenomena, including the search for the EM counterpart to GW alerts. {\it Swift} houses three instruments: the Burst Alert Telescope \citep[BAT; 15–350\,keV;][]{bar05}, the X-ray telescope \citep[XRT; 0.2-10\,keV;][]{bur05} and the Ultra-violet/optical Telescope \citep[UVOT; 1600–8000\AA;][]{roming}. The large field of view of the BAT, 1.4 sr (50 per cent coded), enables it to continuously view the sky and alert the spacecraft to new gamma-ray transient events such as GRBs. Once a new transient has been discovered {\it Swift} rapidly slews to enable the two narrow-field instruments to observe the error region. The XRT is a focusing instrument with a peak effective area of $110\,{\rm cm^2}$ at $1.5\,$keV and a roughly circular field of view with radius 11.8\,arcmin. The UVOT has six optical/UV filters covering 1600–6240\,\AA\,and a $white$ filter covering 1600–8000\,\AA, with a peak effective area of $50\, {\rm cm^2}$ in the $u$-band. The UVOT field of view is $17\times17\,{\rm arcmin}$. Chasing the EM counterpart to a GW alert is in general difficult due to the large uncertainty in the location of these events on the sky; the probability regions released by LIGO-Virgo during O3 ranged from tens to thousands of square degrees. {\it Swift} has an advantage in the chase for the EM counterparts as it can respond quickly, commencing observations within a couple of hours of the GW alert, and can observe large portions of the sky within 24 hours \citep{eva16,kli19,pag20}. {\it Swift} has already shown its importance with the detection of the UV counterpart to the GW trigger 170817 by UVOT \citep{eva17}. BNS mergers are expected to be accompanied by a KN; red, thermal emission, produced when the ejected material, rich in neutrons, forms heavy elements through rapid neutron capture (r-process) nucleosynthesis \citep[e.g.,][] {lat74,lip98,met10,bar13,ros14,cho17,cow17,dro17,kas17,mcc17,nic17,sha17,sma17,tan17,uts17,vil17} and subsequently decays radioactively. High-opacity lanthanide-rich ejecta produced during the merger is expected to suppress UV and optical emission. The discovery of the UV counterpart provided the first evidence for a lanthanide-poor wind producing this blue emission \citep{and17,arc17,cho17,cow17,dia17,dro17,eva17,kas17,mcc17,nic17,pia17,sma17,sha17,tan17,val17,vil17}.

EM radiation is expected to be produced for both BNS and BH-NS mergers, but the emission characteristics depend on the geometry of the system. If the viewer lies close to the axis of rotation, then we expect to observe a short GRB \citep{eic89,nar92}. The KN component is expected to be more isotropic \citep[observed over all angles;][]{lip98,met10}. On the other hand, BBH mergers are not typically expected to produce EM radiation \citep{met19}, but there have been predictions of EM radiation under certain circumstances, such as if accreting circumstellar material is present \citep[see][and references therein]{per19}. \cite{gra20} proposed the discovery of an optical EM counterpart of a BBH merger, GW190521g \citep{abb20e}. For this event, the EM emission is thought to originate from the kicked BBH merger in the accretion disk of an active galactic nucleus (AGN).

There is a large diversity expected in the observed KN \citep{lip98,met19,kaw20}. The KN emission depends strongly on the properties of the ejecta \citep[e.g. mass, density and composition;][]{jus15}, which in turn depend on the properties of the binary components \citep[the type of merger, their masses and spins;][]{met18,met19,kaw20}. The fate of the system post merger also strongly affects the expected KN emission. Different KNe are expected in the BNS scenarios where the merger directly collapses to a BH, has an intermediate phase as a super/hypermassive NS or leaves a stable NS. In the BH-NS scenario, different KNe are expected if the NS is swallowed whole or tidally disrupted \citep{kaw20}. The observer's viewing angle may also affect the colours and luminosity of the observed KN emission \citep{wol18,met19,kor20}\citep{wol18,met19,kor20}. Magnetic fields may also affect the observed emission, for instance, by enhancing winds producing the blue emission \citep{met18} or through magnetic spin-down of a highly magnetised NS \citep{met14b}. 

As AT 2017gfo/GRB 170817A is the only secure detection of an EM counterpart to a CBC trigger so far, each additional detection is important to further our understanding of CBCs. Studies of additional events will be crucial in gaining clear constraints on the actual range in behaviour/properties. {\it Swift}/UVOT is the only instrument that can provide prompt UV observations, which is critical in forming a complete picture of the EM emission associated with a GW event. 

In this paper, we discuss the {\it Swift}/UVOT GW pipeline and the follow-up of GW alerts during O3. One effect of scanning vast areas of the sky for the EM counterpart is discovering a multitude of transient phenomena that are not necessarily related to the GW itself, and we summarise these optical transients found serendipitously during O3 in this paper. For a corresponding analysis on the X-ray observations, we refer the reader to the companion paper by \cite{pag20} which presents the corresponding X-ray data from {\it Swift}. For details on the CBCs GW events observed by LIGO and Virgo during O3a (the first half of the O3 period), see the Gravitational-Wave Transient Catalog \citep[GWTC-2;][]{abb20d}. In \S~\ref{swiftGW} we briefly review the {\it Swift} strategy for follow-up of GW alerts and give an overview of the UVOT candidate identification pipeline. In \S~\ref{SwiftGWSummary} we provide a summary of the {\it Swift}/UVOT GW follow-up effort during the O3 period. Finally, in \S~\ref{discussion} we discuss the sources of interest found by {\it Swift}/UVOT during the follow-up of GW alerts in the O3 period and discuss the importance of {\it Swift}/UVOT in the EM follow-up of GW events.  We conclude in \S~\ref{conclusions}. All uncertainties throughout this paper are quoted at 1$\sigma$ unless otherwise stated. Throughout, we assume the Hubble parameter $H_0 = 70\;{\rm km s}^{-1}\;{\rm Mpc}^{-1}$ and density parameters $\Omega_{\Lambda}= 0.7$ and $\Omega_m= 0.3$. All magnitudes are given in the AB system unless otherwise stated.

\section{{\it Swift} and Follow-up of Gravitational Wave Sources}
\label{swiftGW}
In the ideal scenario, when GWs are emitted by merging objects, a short GRB will also be produced, which triggers the usual response by {\it Swift}. For this to occur, the merger should be a BNS or BH-NS, and the Earth should lie along or close to the rotation angle of the merger. However, the opening angles of the jets are expected to be narrow, between $3-8\deg$ \citep{bur06,fon15,tro16,jin18}. If the jet is characterized by an angular structure, as seen in GW170817 \citep[e.g][]{hag17,hal17,mar17,ale18,lym18,moo18,fon19,ghi19,haj19,lam19,tro19,tro20}, then its prompt emission could be detected for even larger viewing angles. The chance of {\it Swift}/BAT detecting the $\gamma$-ray emission from the short GRB resulting from the merger of a BNS or BH-NS detected by LIGO-Virgo is still very small \citep[$\approx 0.2\,{\rm yr^{-1}}$;][]{dic20}. This rate becomes approximately 3 times higher with Gamma-ray Urgent Archiver for Novel Opportunities (GUANO) targeted searches \citep[see][and DeLaunay and Tohuvavohu in prep.]{toh20}. Fortunately, the scheduling of {\it Swift} is highly flexible and responsive. Once a GW alert has been received, it can respond in a matter of hours to cover substantial portions of the GW error region with the XRT and UVOT in relatively short amounts of time in order to detect any accompanying X-ray and optical/UV emission \citep[see][]{eva16,kli19,pag20}.

\subsection{{\it Swift} Observing Strategy}
The longer-lived EM counterpart to a GW alert (expected to be associated with a BNS or BH-NS trigger, e.g. a KN or GRB) is likely to be produced at X-ray and longer wavelengths. Therefore the {\it Swift} observing strategy is optimised for follow-up with the XRT and UVOT instruments. This strategy was described in detail by \citet{eva16b,eva16}, but we provide a summary here. Since the XRT and UVOT have narrow fields of view and the error region of a GW alert may cover many square degrees, many pointings (tiles) must be used to cover even a fraction of the probability region. Even with tiling, the majority of the error circle may not be covered within a reasonable time. Therefore it is essential to try and place further constraints to prioritise regions of the sky within the GW error region. Since CBCs are expected to occur in or near galaxies, a reasonable strategy is to convolve the LVC probability region and the distance of the GW trigger with a galaxy catalogue \citep[see \S~3.2 of][for further details]{eva16,eva19}. We use one of two catalogues: the 2MASS Photometric Redshift catalogue \citep[2MPZ;][]{bil14} or the Gravitational Wave Galaxy Catalogue \citep[GWGC;][]{whi11}. The GWGC catalogue is more complete compared to the 2MPZ for nearby distances \citep[for further discussion see][]{eva16}. Therefore we convolve the LVC probability region and the distance of the GW trigger with GWGC if the GW event is $<80$ Mpc and with 2MPZ if the GW event is $>80$ Mpc. Historically, the observing strategy was optimised to enable the greatest coverage with the XRT. This meant that some of the probability region, potentially containing host galaxies, would not fall within the UVOT field of view. However, UVOT proved to be an important discovery instrument with the UV detection of AT 2017gfo \citep{eva17}. Therefore a change was made to the observing strategy before the start of O3. The strategy was adjusted so that galaxies with a high probability of being the host will fall entirely within the UVOT field of view \citep{kli19}.

\subsubsection{Strategy for activating {\it Swift} follow-up of GW events during O3}
\label{trigger_strategy}
Once a convolved probability map\footnote{The convolved map is created from the LIGO-Virgo probability map and the galaxy catalogue.} is created, a decision can be made on whether to follow up a given GW alert. This is based on the trigger type, the FAR, and how much of the error region {\it Swift} can cover within one day. This essentially implies that {\it Swift} observes, primarily, events that are well localised and have a high chance of producing EM radiation, e.g. BNS merger \citep{eva16,eva16b,kli19}. This strategy implicitly includes distance because the area to be tiled for nearby events will be reduced due to the galaxy convolution. Midway through the O3 period, the triggering strategy was modified to consider the probability that the trigger could have a terrestrial origin and to place stronger weight on those events where a NS is likely to have been disrupted and therefore likely to produce an EM counterpart. We summarise this updated strategy in the rest of this section. For a comparison of the strategy implemented in the O3a and O3b phases, see \cite{pag20}. 

Unmodelled triggers do not require a well-known or accurate waveform model. Unmodelled triggers may be a range of transients: CBCs, core-collapse supernovae (CCSN), neutron star quakes and other phenomena that may be more exotic such as cosmic strings \citep{abb09,lyn17}. For unmodelled triggers for which the GW central frequency is $>$1Hz, these events may be Galactic in origin\footnote{Since the power needed to create a GW signal goes as the sqaure of the frequency, then to generate a high-frequency signal the object must either be nearby or have a lot of mass from which to generate the GW radiation strong enough to trigger LIGO-Virgo. However, with a larger mass we generally get lower frequencies (i.e. orbital periods are longer). Therefore the most likely source of a high-frequency trigger is a very nearby stellar object, i.e. Galactic (Veitch, private communication).} and we follow all events. If the GW signal of the unmodelled trigger is $<$ 1Hz, these are unknown, perhaps exotic objects. These events are only followed if we expect to obtain reasonable coverage, or there has been an announcement of the detection of a counterpart. 

The strategy for follow up of CBC triggers is split depending on the probability of containing a disrupted NS (DNS). The probability is determined using the following:
${\rm P_{DNS}} = {\rm P_{NS}}*(1-{\rm P_{TERRES}})-{\rm P_{BHNS}} +
{\rm P_{REMNANT}}* {\rm P_{BHNS}}$
where $\rm {P_{NS}}$ is the LVC probability that the event contains a NS, ${\rm P_{TERRES}}$ is the probability that the event is terrestrial in origin (e.g. noise) and ${\rm P_{BHNS}}$ is the probability that the event is a BH-NS merger. For each trigger, ${\rm P_{NS}}$, ${\rm P_{TERRES}}$, ${\rm P_{BHNS}}$ and ${\rm P_{REMNANT}}$ are taken from the relevant GCN notice sent by the LVC. For BH-NS triggers, the ${\rm P_{REMNANT}}$ field indicates how likely a remnant is, so we include the chance, ${\rm P_{REMNANT}}* {\rm P_{BHNS}}$. In this instance ${\rm P_{DNS}} = 1$ implies a NS was disrupted and 0 implies no NS was disrupted. 
The decision tree was set the following way: {\it Swift} would
follow up an event only in the following cases:\\
{\it 1. For burst (unmodelled) triggers:}\\
(a) if $\geq 1$\,kHz;\\
or (b) if $<1$\,kHz and FAR $<1/$year and well localised (50 per cent of the probability, post galaxy convolution, is observable by XRT
within 24 hours).\\
{\it 2. For CBC triggers:}\\
(a) if ${\rm P_{DNS}} = 0$, FAR $< 1/10$ years and 50 per cent of the probability, post galaxy convolution, is observable by XRT
within 24 hours;\\
or (b) if $0 < {\rm P_{DNS}} \leq 0.25$, FAR $< 1/10$ years and 50 per cent of the probability, post galaxy convolution, is observable by XRT within 24 hours;\\
or (c) if $0.25 < {\rm P_{DNS}} \leq 0.7$ and $>40$ per cent of the probability, post galaxy convolution, is observable by XRT within 24 hours;\\
or (d) if ${\rm P_{DNS}}> 0.7$ and $>10$ per cent of the probability, post galaxy convolution, is observable by XRT
within 24 hours.\\

These thresholds are set to optimise the balance between the time {\it Swift} dedicates to GW follow up and the potential science return. CBC trigger types, (a) and (b) typically will correspond to a BBH merger or BH-NS whereby the NS is not disrupted and may have been swallowed whole. In these cases, the expectation of observing an EM counterpart is low, but the science return is high if any emission were to be detected \citep[e.g][]{shi19,met19,mck19,per19}. In the cases where no EM emission is detected, but there is good coverage of the probability region, useful constraints can be placed on the expected emission. CBC trigger types, (c) and (d) will typically correspond to a BNS or BH-NS merger whereby the NS is likely disrupted. In these cases, the expectation of observing an EM counterpart is higher. It is worth following these events even if only a low percentage of the post galaxy convolution is observable by XRT within 24 hours.

\subsubsection{{\it Swift} strategy for observing GW events}
Once a decision has been made to activate {\it Swift} follow-up of a GW event, the appropriate observational strategy is decided. This strategy remained the same for all trigger types until the midpoint of O3 \citep[see][]{eva16,kli19,pag20}. For the different cases of unmodelled and CBC triggers described above (\S\ref{trigger_strategy}), the strategy after the O3 midpoint is as follows:\\
{\it 1. For burst (unmodelled) triggers:}\\
(a) Convolve LVC probability map with Galactic plane and then observe fields for 80\,s each\footnote{The actual exposure time is slightly less due to the spacecraft slewing to the target, and ramp up of the photocathode in UVOT.}. Once complete, re-observe all fields for
80\,s. Keep repeating, only stop when a counterpart is found or four days have passed. \\
(b) Observe 800 fields or 90 per cent of the galaxy-convolved probability (whichever is smaller) for 80\,s each. If possible, repeat observations in the same field for up to 3 days. Then, observe 500\,s per field until all fields are re-observed or four days of these observations have been completed.\\
{\it 2. For CBC triggers:}\\
(a) Follow for 24hrs. If 90 per cent of the (post galaxy convolution) probability can be observed in 48 hours, then follow for 48 hours.\\
(b) Observe 500 s per field for four days or until 90 per cent of the probability has been covered (whichever comes first). Do not start until 12 hours after the LVC trigger time.\\
(c) and (d) the same strategy as for burst (unmodelled) type (b) triggers. \\

The CBC (b) mergers have a low probability of containing a disrupted NS. These triggers are likely to be BBH mergers. BBH mergers are not expected to produce EM emission immediately after the trigger (if at all), and so the start of observations is delayed by 12hrs. The majority of observations performed by the UVOT are with the $u$ filter. This filter has the largest throughput after the $white$ and $b$ filters but is bluer than is typically performed by ground-based telescopes. In around 10 per cent of tiles, a less sensitive filter (e.g. $uvw1$) or the blocked filter is used to avoid damage to the instrument due to bright stars/fields.

\subsubsection{{\it Swift}/UVOT Archival Coverage}
\label{SGWGS}
The {\it Swift}/UVOT archive covers 15.3 per cent of the sky. In the $u$-band, observations cover 8.4 per cent of the sky. Due to the vast error region of the LIGO-Virgo triggers, the likelihood of there being an archival UVOT image is small. There is a concerted effort to build an archive of local galaxies out to 100 Mpc with UVOT to have a template comparison image in the event of a nearby GW alert. The {\it Swift} Gravitational Wave Galaxy Survey (SGWGS; Tohuvavohu et al., in prep) is presently observing 4773 fields, which comprise 13000 galaxies. The survey will cover 41.8 per cent of all the catalogued B-band luminosity within 100 Mpc. However, the survey is optimised for the XRT, which has a larger field of view than the UVOT. Therefore only 76 per cent of the galaxies are expected to be within the UVOT field of view. Once complete, it is expected that per GW alert with $60-100$ deg$^2$ error region, on average, the brightest ten galaxies in the error region will have archival images for comparison. 

\subsection{UVOT GW pipeline}
The UVOT GW pipeline uses the standard output files produced by the processing pipeline at the {\it Swift} Data Centre (SDC)\footnote{https://swift.gsfc.nasa.gov/sdc/}, which processes all {\it Swift} data. The UVOT GW processing pipeline is alerted when a new tiling observation has been processed by the SDC and is available at the SDC Quick Look web site\footnote{https://swift.gsfc.nasa.gov/sdc/ql/}. Data from that tile (a sequence) is downloaded to a Unix workstation at GSFC's Astrophysics Science Division, and the UVOT sky images are searched to identify new transient sources that might be the counterpart to the GW event. For each observation, the ``best" UVOT image is chosen (for the vast majority of tilings during the O3 period UVOT produced a single $u$-band exposure per tile) and the ftool \textsc{uvotdetect} \citep[based on \textsc{Sextractor} ][]{ber96} is run for that exposure. All the sources found are run through a series of checks to determine if they are likely to be previously known sources, extended sources, or sources due to image artefacts. The checks include: comparing the position to that of sources in the USNO-B1.0 Catalog \citep{mon03} and the HST Guide Star Catalog II \citep{las08}, comparing the size of the major and minor axes of the source with that expected for a point source, and nearness to other bright UVOT sources. Sources due to image artefacts are avoided by comparing the position to those expected for readout streaks, and smoke rings \citep{bre11,pag14}. Sources that pass all these initial tests are then further checked against the Gaia Catalog DR2 \citep{gai18} and the Minor Planet Checker\footnote{https://cgi.minorplanetcent.net/cgi-bin/checkmp.cgi}. Every source found by \textsc{uvotdetect} is assigned a Quality Flag based on the results of these checks. Sources deemed more likely to be real transients are assigned lower numbers; sources that pass all the checks are assigned a flag of 0 or 1 depending on their magnitude, referred to as Q0 or Q1 sources, respectively. Sources dimmer than a magnitude of 19.9 (a conservative sensitivity limit to obtain a signal to noise $>$5 in the $\sim 80$s tiling observations) are assigned a value of 1.

The pipeline reliably finds new sources if they are isolated from existing sources and not affected by the defects in the UVOT images caused by very bright sources. However, it also produces some false detections because of the large number of UVOT sources that have to be evaluated. Consequently, the pipeline produces small images (thumbnails) for all Q0 and Q1 sources. The thumbnails allow scientists to evaluate the reliability of possible UVOT counterparts quickly. Each thumbnail has an associated flag giving the quality rating or identifying why the thumbnail was produced. The complete set of thumbnail identifiers are described in Table \ref{flag_description}. Since the source-finding software (\textsc{uvotdetect}) sometimes misses new sources within extended sources such as nearby galaxies, thumbnails are produced for nearby galaxies reported in the GLADE catalogue \citep[v2.3;][]{dal18}\footnote{The GLADE catalogue was favoured as it was more complete than other galaxy catalogues, see Fig 4. \cite{dal18}.} that are observed with UVOT. Thumbnails are also produced for XRT counterparts categorised as rank 1 or rank 2 sources\footnote{Rank 1 and rank 2 sources meet the criteria if they are uncatalogued and at least 5$\sigma$ and $3\sigma$, respectively above the $3\sigma$ upper limit from RASS or 1SXPS; or a known X-ray source which is $5\sigma$ or $3\sigma$ above the catalogued flux.}. Also, if a Q0 source has an archival {\it Swift}/UVOT SGWGS image, a thumbnail with the quality flag `2uvot' is produced, enabling like for like comparison. Thumbnails were, however, not produced automatically for other archival UVOT images but were downloaded and examined during manual inspection of candidate sources. Thumbnails were also not created for sources flagged as Q2 or Q3. These are considered to be known sources.

\begin{table}
\centering 
\caption{Description of the flags given to thumbnails images created for individual UVOT and XRT sources.}
\begin{threeparttable}
\begin{tabular}{cp{6cm}} 
\hline 
Quality Flag & Description \\
\hline
Q0 & a UVOT source that passes all the quality checks and is brighter than 19.9 mag \\
Q1 & a UVOT source that passes all the quality checks but is fainter than 19.9 mag. There is no magnitude limit for Q1 sources as long as uvotdetect finds them$^\dagger$. \\
Q2 & a Q0 source but with a match to a catalogued object. Matching parameters: within an angular distance of $2.5\arcsec$ and catalogue magnitude within 2 magnitudes of the UVOT object.\\
Q3 & a Q1 source but with a match to a catalogued object. Matching parameters: within an angular distance of $2.5\arcsec$ and catalogue magnitude within 2 magnitudes of the UVOT object.\\
\hline 
Other Flag & Description \\
\hline
2uvot & Q0 UVOT sources with images of current and archival UVOT exposures if available \\
gal & GLADE galaxies detected in UVOT \\
xrt & XRT counterparts flagged as rank 1 or rank 2 located with a UVOT image \\
\hline 
\hline 
\label{flag_description}
\end{tabular} 
  \begin{tablenotes}
    \item[$\dagger$] \textsc{uvotdetect} uses a threshold of 2.0 standard deviations above the noise
  \end{tablenotes}
\end{threeparttable}
\end{table}

The UVOT pipeline was modified and improved several times during O3 to reduce the number of thumbnails to check and avoid missing faint sources that may be the GW counterpart. Thumbnails for Q1 sources were added from mid-July 2019 onward. At the same time, thumbnails for $uvw1$ images ceased to be produced as these images were found to contain a high number of false sources.

For a typical {\it Swift} follow-up of a GW error region covering tens of deg$^2$, the pipeline produces on average 2000 to 3000 thumbnails. The majority are thumbnails of galaxies, identified by the quality flag `gal'. Approximately 100-200 thumbnails have other quality flags such as Q0, Q1. During O3, UVOT performed 6441 observations and the pipeline created 18459 thumbnails.

\subsubsection{Candidate Inspection}
\begin{figure*}
\includegraphics[angle=0,scale=0.65,trim={7.0cm 0.0cm 0.cm 0.0cm},clip]{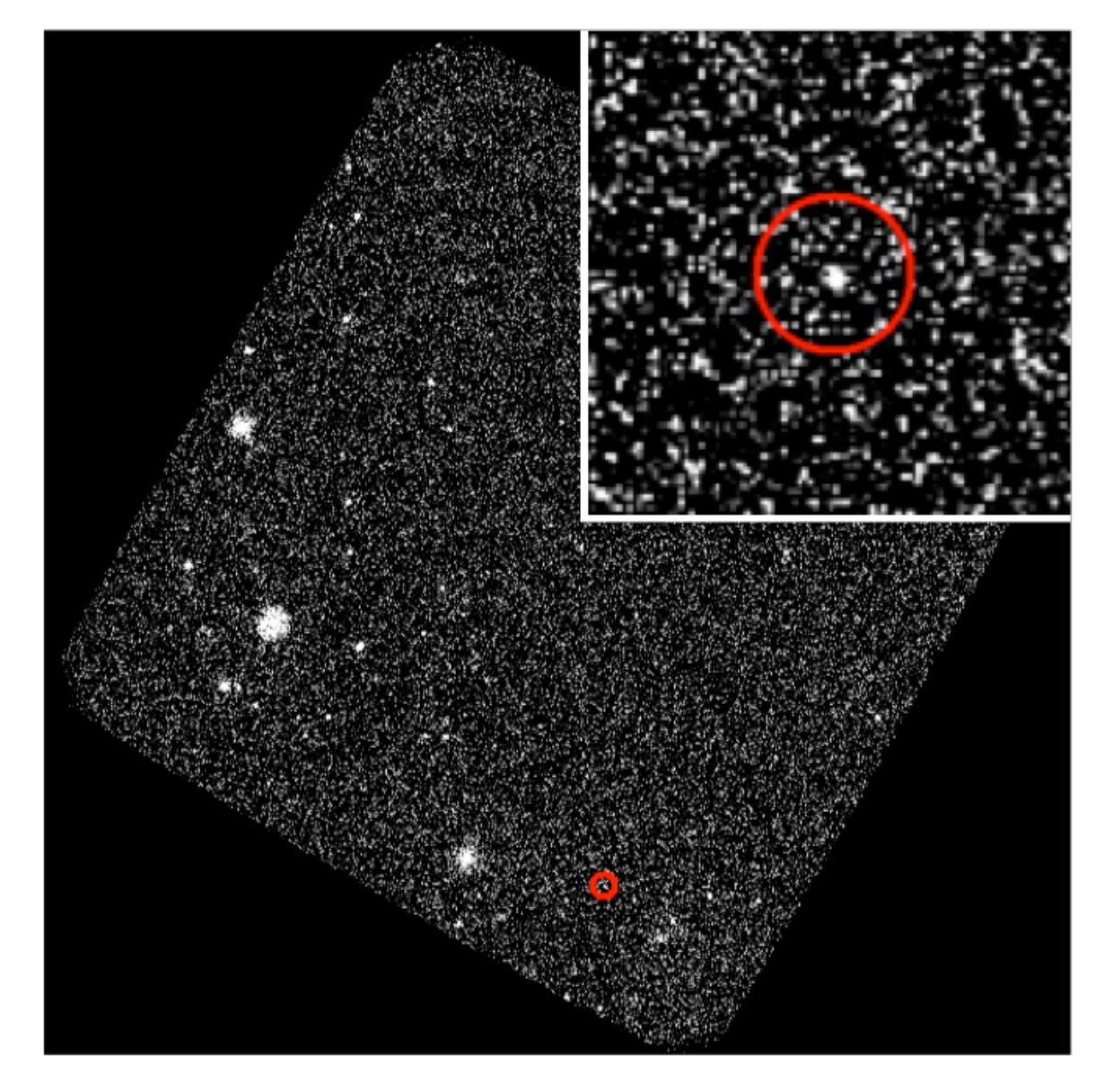}
\includegraphics[angle=0,scale=0.65,trim={7.0cm 0.cm 0.cm 0cm},clip]{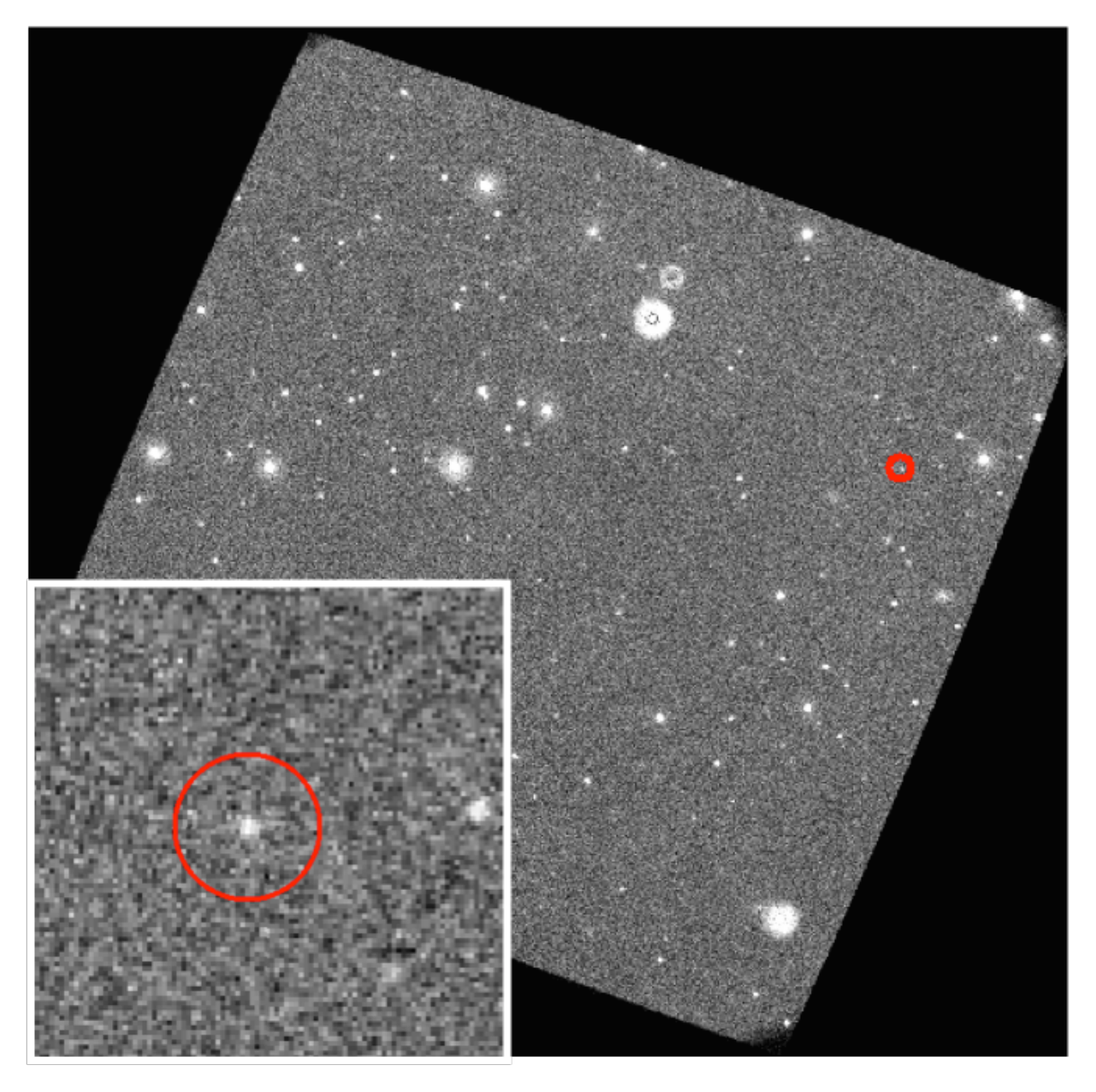}
\includegraphics[angle=0,scale=0.65,trim={7.0cm 0.0cm 0.cm 0cm},clip]{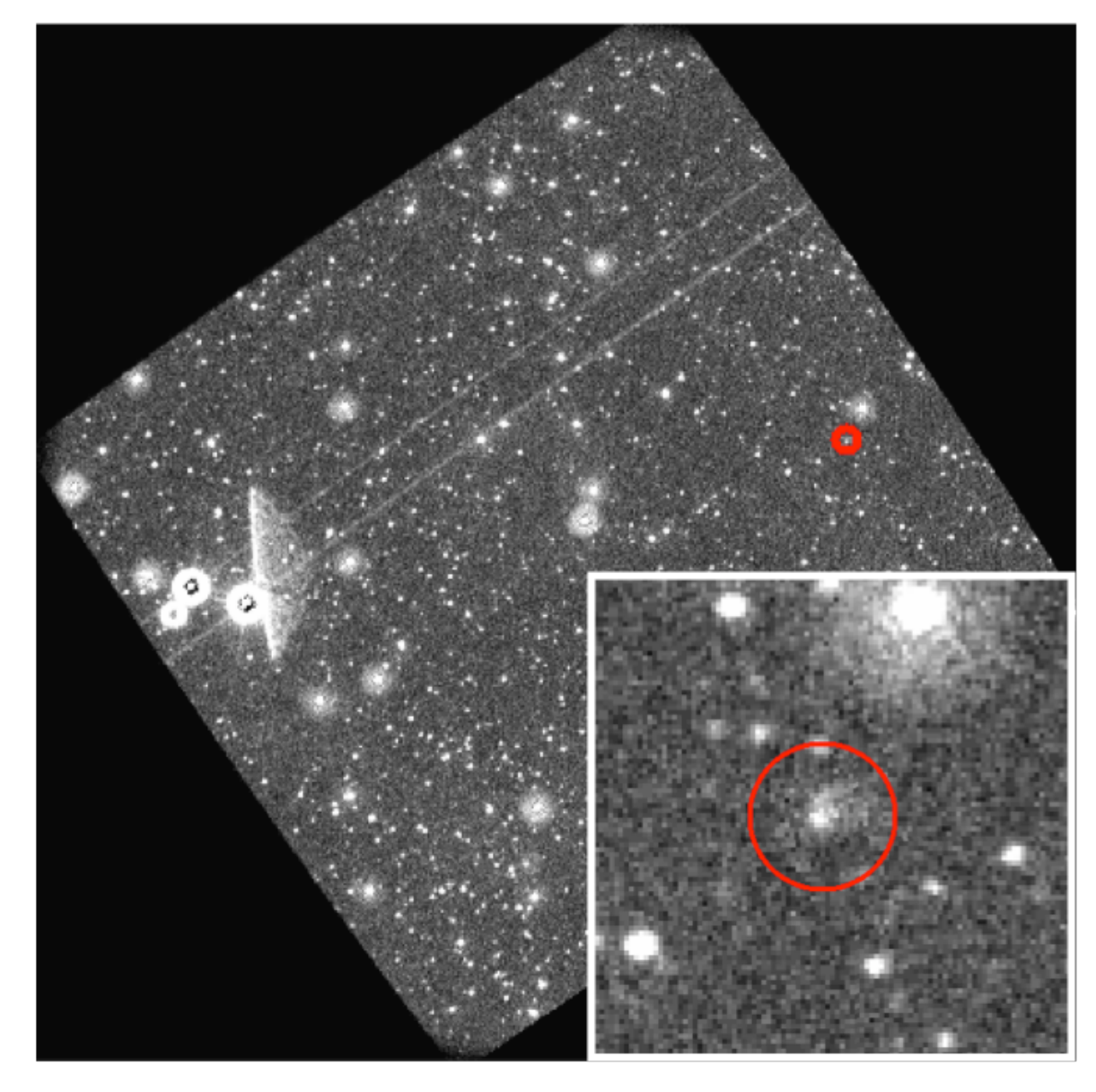}
\includegraphics[angle=0,scale=0.65,trim={7cm 0cm 0.cm 0cm},clip]{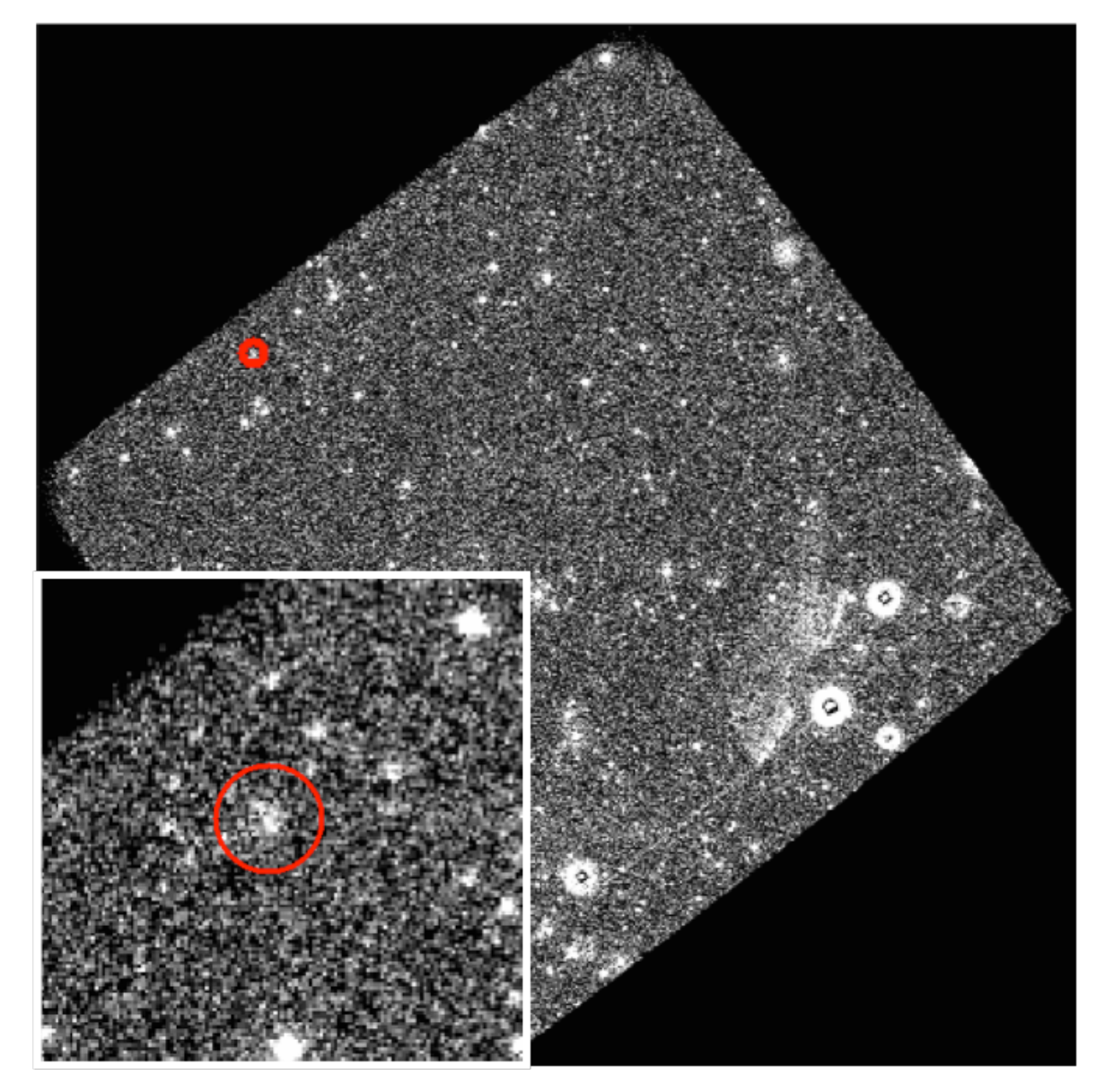}
\caption{Examples of Q0 (left) and Q1 (right) sources. The top panels are identified as confirmed astrophysical sources, while the bottom panels are ghosts. The ghosts are scattered light features that result from secondary reflections within UVOT when there is a bright source in the field of view. The red circle indicates the location of the Q0 or Q1 source on the image. A zoom-in is provided in the insert. For the ghosts, the origin of this scattered light feature is likely the brightest source on the opposite side of the image.}
\label{scattered_light_ghost}
\end{figure*}

The thumbnails produced by the pipeline for each tile must be visually inspected to verify candidate counterparts before they are released to the community. Candidate inspection is important because scattered light artefacts \citep{pag14}, which are inherently difficult to predict, may be misidentified as Q0 or Q1 sources. Those that are identified by eye as due to scattered light are rejected. We have newly identified a rare scattered light artefact through the manual inspection of candidates, which we label as `ghost'. These sources are small (a few arcsec in diameter) point-like or smudge-like sources that appear on images where there is a bright source in the field of view, which produces strong scattered light features. The ghosts are likely a result of secondary reflections within the instrument. Ghosts are not expected to be produced by bright stars outside the UVOT field of view. Stars outside the FOV are instead expected to produce streaks, as observed in {\it XMM-Newton}-OM, but this is mitigated in UVOT by the housing. In Fig. \ref{scattered_light_ghost} we show examples of Q0 and Q1 sources in both short ($\sim 80$\,s) and long ($\sim 500$\,s) exposures, identified as astrophysical sources and ghosts. In the examples of images containing ghosts, the ghosts are more diffuse than the neighbouring astrophysical sources and are less bright than sources of a similar dimension. Since only a handful of these artefacts have been identified, we are not yet able to automatically exclude these sources, and as such, these need to be manually rejected. A check is also made for any nearby high proper motion sources to ensure that the candidate is a new source and not an existing known source that has moved.

To look for changes in brightness in the galaxy thumbnails, difference imaging should ideally be performed. However, for the UVOT, there are only a small number of archival $u$-band UVOT images. Therefore, the thumbnails of galaxies are manually scrutinised for changes in brightness or any new point sources by manually comparing with the archival UVOT image if available or the DSS image. 

The positions of any candidates remaining after these initial checks are cross-checked against additional archival catalogues and images \citep[e.g. using the VizieR facility at the CDS; ][]{och00}, including checks against the GALEX archive \citep{bia14} to determine if the source is of astrophysical interest. Sources that are of immediate interest to the astronomical community, once manually vetted, are released through the GCN network. In this paper, we summarise all sources of interest that, upon manual inspection, are new sources or have a $3\sigma$ increase in brightness compared to historical values. Some of these sources were deemed to be of immediate interest to the GW-EM community and were reported via GCN Circulars.

\subsection{UVOT data analysis}
\label{uvotdataanalysis}
All images were downloaded from the {\it Swift} data archive\footnote{https://www.swift.ac.uk/archive/index.php}. To determine the magnitude of a source, we use the following method. If the source is below 0.5 counts/s (c/s) and its position is known, either because it was identified by another facility or a known catalogued UV/optical/nIR source was present, we used a circular region with a $3\arcsec$ radius. In all other instances we use a $5\arcsec$ radius. In order to be consistent with the UVOT calibration \citep{poole}, count rates extracted with a $3\arcsec$ region were then corrected to $5\arcsec$ using the curve of growth contained in the {\it Swift} calibration files\footnote{https://heasarc.gsfc.nasa.gov/docs/heasarc/caldb/swift/}. Background counts were extracted using an annular region of inner radius $15\arcsec$ and outer radius of $30\arcsec$, or an aperture with a comparable area, from a blank area of sky near to the source position. The count rates were obtained from the image lists using the {\it Swift} tool {\textsc {uvotsource}}. Finally, the count rates were converted to AB magnitudes using the UVOT photometric zero-points \citep{bre11}. The analysis pipeline used UVOT calibration 20201008. The UVOT detector is less sensitive in a few small patches\footnote{https://heasarc.gsfc.nasa.gov/docs/heasarc/caldb/swift/docs/
\\uvot/uvotcaldb\_sss\_01.pdf} for which a correction has not yet been determined. Therefore, we have checked to see if any of the sources of interest fall on any of these patches in any of our images and exclude five individual UV exposures (2 for ZTF19aarykkb and 3 for ZTF19acymixu) for this reason. For sources contaminated by an underlying galaxy, we have provided background subtracted values for those which have template observations available or for which an estimate of the flux could be obtained from an offset position on the host galaxy. The details of host subtraction are provided for individual events in the supplementary section.

\section{Results of {\it Swift}/UVOT follow-up during the O3 period}
\label{SwiftGWSummary}
Of the 56 public GW alerts released by the LVC, {\it Swift} obtained images for 18 GW alerts. These events were selected as they met the $\it Swift$ trigger criteria, as given in \S \ref{trigger_strategy}. However, two of the events were followed at the request of the {\it Swift} team for reasons explained below. Of these 18 events, three were retracted by the LVC, and a further three of the 18 GW alerts did not match the criteria for the Gravitational-Wave Transient Catalog (GWTC-2) of CBCs observed by LIGO and Virgo \citep{abb20d}. The fraction of the LVC region tiled by UVOT varies from alert to alert. We summarise our search of all GW alerts which had a partial or complete response in this section. A list of GW alerts followed by {\it Swift} along with their classification, distance, FAR, the fraction of LVC region covered by the UVOT, the number of fields observed by UVOT and the number of thumbnails produced during automatic analysis by the UVOT software is given in Table \ref{tab:obs}. Full details of the fields observed in each case can be found under the specific trigger page at https://www.swift.ac.uk/LVC/. For the complete list of GW alerts from the third observing run, noting which were followed-up by {\it Swift}, see Supplementary A1 of \cite{pag20}. Detailed information about the {\it Swift} pointings is also provided by The Gravitational Wave Treasure Map tool\footnote{http://treasuremap.space/} \citep{wya20}, which is designed to visualise and coordinate EM follow-up.

\subsection{GW alerts followed-up with {\it Swift}/UVOT}
In the following, we provide a summary of the individual GW triggers followed-up by {\it Swift} during the O3 period and provide an overview of the results of the processing and inspection of UVOT images and thumbnails produced during the automatic pipeline processing. We provide the number of sources of interest determined, after manual inspection, to be astrophysical in origin and to have brightened by 3$\sigma$ compared to archival values. We also provide a summary of the {\it Swift}/UVOT follow-up of sources discovered by other observatories. Detailed information on the individual sources, their detection and follow-up are provided in Supplementary S.1. 

For candidate sources reported by other observatories, two names may be given: the name given by the instrument team and the name given by the Transient Name Server\footnote{https://wis-tns.weizmann.ac.il/} (TNS), e.g. AT2019aaa. We summarise all sources of interest identified or followed-up by UVOT in Table \ref{tab:UVOTsrcsmmry}. We provide the UVOT photometry in Table \ref{tab:UVOTobs} and give the XRT count rate for each source in Table \ref{tab:XRTsrcsmmry}. We note that none of the sources of interest were detected by XRT in either single exposures or a stack of all X-ray observations taken at that location.

\begin{landscape}
\begin{table}
\caption{The GW alerts triggering a full or partial response by {\it Swift} together with statistics on the UVOT response. The columns given are the LVC GW trigger name, the {\it Swift} identification number (ID), the GW trigger time (${\rm T_{0,GW}}$) in UT, the start time of the {\it Swift} observations in hours after trigger time, the false alarm probability per year (FAR), the sky localization (area; 90 per cent probability region), the distance to the GW and 1$\sigma$ error, the number of observations performed by UVOT, the fraction of the total localization probability tiled by UVOT, the number of thumbnails produced in the UVOT GW pipeline, the GW classification with the chance of it being a particular type of object given as a percentage in brackets. In the FAR and area columns and table notes, we identify which skymap these values were taken from BAYESTAR, LALInference or coherent WaveBurst respectively, see https://gracedb.ligo.org/. A large FAR value indicates a low likelihood of this event being a false event.}
\label{tab:obs}
\begin{threeparttable}
\begin{adjustbox}{width=23cm}
\begin{tabular}{ccccccccccccc}
\hline 
GW trigger &  {\it Swift} ID & ${\rm T_{0,GW}}$ (UT) & {\it Swift} obs. & FAR (${\rm yr^{-1}}$)\tnote{a} & Area\tnote{a}  & Distance\tnote{a}  & Number of & Fraction of GW  & Unique Area  & Number of & GW Classification\tnote{a} \\
    	   &       	     &           	     & start time       & 	 		         &   90 per cent   &  (Mpc)        &  UVOT  & localization probability  & Tiled by     & thumbnails &  \\
    	   &      	     &           	     &   (hrs)          &			         &  (deg$^2$)  &            &  Observations           &   covered by UVOT  & UVOT (deg$^2$)  &  &   \\
\hline	 
S190412m    & 19 & 05:30:44 & $T_0$+10.6 & $1.88\times10^{19}$ &    156\tnote{b} & $ 812\pm194$\tnote{b} & 100 & 0.12  & 6.59 & 287    & BBH ($>99$)  \\				
S190425z    & 22 & 08:18:05 & $T_0$+4.6  & $6.98\times10^{4}$  &   7461          & $ 156\pm 41$          & 406    & 0.01  & 29.22 & 2773   & BNS ($>99$)  \\ 
S190426c    & 23 & 15:21:55 & $T_0$+2.4  & 1.63                &   1131          & $ 377\pm100$          & 894    & 0.13  & 43.77 & 1008   & BNS (24), MassGap (12), \\
            &    &          & $   $      &                     &                 &                       &        &       &          &        & Terrestrial (58), BH-NS (6), BBH ($<1$)\\ 
S190510g    & 25 & 02:59:39 & $T_0$+2.0  & 3.59                &   1166          & $ 227\pm 92$          & 977    & 0.46  & 55.71 & 2836   & Terrestrial (58), BNS (42) \\ 
S190718y    & 39 & 14:35:12 & $T_0$+3.8  & 1.15                &   7246\tnote{b} & $ 227\pm165$\tnote{b} & 368    & 0.12  & 20.88 & 734    & Terrestrial (98), BNS (2) \\
S190728q    & 42 & 06:45:11 & $T_0$+12.7 & $1.25\times10^{15}$ &   104           & $ 874\pm171$          & 144    & 0.22  & 11.23  & 115    & BBH (95), MassGap (5) \\
S190808ae$^{\rm R}$ & 43 & 22:21:21 & $T_0$+3.4  & 1.06        &   5365\tnote{b} & $ 208\pm 77$\tnote{b} &  36    & $<$0.01 & 2.91        & 365    & Terrestrial (57), BNS (43) \\	
S190814bv   & 44 & 21:10:39 & $T_0$+3.2  & $1.55\times10^{25}$ &     23          & $ 267\pm 52$          & 352    & 0.69  & 14.83 & 2488   & BH-NS ($>99$) \\
S190822c$^{\rm R}$  & 46 & 01:29:59 & $T_0$+2.0  & $5.16\times10^{9}$  &   2769\tnote{b} & $  35\pm 10$ 	 & 37     & $<$0.01 & 2.91    & -      & BH-NS ($>99$), Terrestrial ($<1$) \\		        
S190930t    & 57 & 14:34:08 & $T_0$+2.1  & 2.05                &  24220\tnote{b} & $ 108\pm 38$\tnote{b} & 746    & 0.02  & 50.06 & 2479   & BH-NS (74), Terrestrial (26)\\
S191110af$^{\rm R}$ & 61 & 23:06:44 & $T_0$+2.9  & 12.68	       &   1261\tnote{c} & $-$          & 797    & $<$0.01 & 87.81   & 55     & Unmodelled transient candidate\\ 	  
S191213g    & 70 & 04:34:08 & $T_0$+39.7 & 1.12                &   4480          & $ 201\pm 81$          & 4   & $<$0.01 & -         & 3      & BNS (77), Terrestrial (23) \\  
S191216ap   & 73 & 21:33:38 & $T_0$+6.1  & $2.80\times10^{15}$ &    253          & $ 376\pm 70$          & 113    & 0.02 & 8.01  & 190    & BBH (99)\\	
S200114f    & 82 & 02:08:18 & $T_0$+1.8  & 25.84               &    403\tnote{c} & $-$          & 206    & 0.22 & 15.53  & 183    & Unmodelled transient candidate \\	
S200115j    & 83 & 04:23:10 & $T_0$+2.0  & 1513                &    765          & $ 340\pm 79$          & 512    & 0.03 & 24.08  & 3266   & Mass Gap ($>99$) \\
S200213t    & 88 & 04:10:40 & $T_0$+5.7  & 1.97                &   2326          & $ 201\pm 80$          & 9      & $<$0.01& 0.62   & 5       & BNS (63), Terrestrial (37) \\
S200224ca   & 90 & 22:22:34 & $T_0$+6.1  & 1974                &     72          & $1575\pm322$          & 670    & 0.62 & 46.17  & 1654	  & BBH ($>99$) \\			
S200225q    & 91 & 06:04:21 & $T_0$+47.9 & 3.45                &     22          & $ 995\pm188$          & 70     & 0.39 & 2.85  & 18	  & BBH (96), Terrestrial (4)\\
\hline 
\hline
\end{tabular}
\end{adjustbox}
  \begin{tablenotes}
    \item[a] Note these parameters are from the most recent analysis on https://gracedb.ligo.org/superevents/public/O3/ and may not necessarily be the values at the time {\it Swift} observations \\
    were uploaded. Area and distance parameters are from LALInference.fits.gz or are otherwise marked.
    \item[b] Value from bayestar.fits.gz 
    \item[c] Value from cWB.fits.gz
    \item[R] Retracted
  \end{tablenotes}
\end{threeparttable}
\end{table}
\end{landscape}

\subsubsection{S190412m/GW190412}
S190412m \citep{GCN:24098} was identified as a BBH ($>99$ per cent). This event formally did not meet the {\it Swift} trigger criteria but was chosen to follow-up as a test of the new {\it Swift} tiling plan (`manypoint') designed to observe many locations in a short space of time. Through further analysis, S190412m was confirmed as a highly significant GW detection and renamed GW190412, a BBH merger with asymmetric masses \citep{abb20}. Automatic analysis identified 13 Q0 sources. Manual inspection of the thumbnails identified 3 of the Q0 thumbnails as being sources of interest.

\subsubsection{S190425z/GW190425}
For S190425z \citep{GCN:24168} a high probability BNS GW alert, automatic analysis identified 22 Q0 sources. All sources except two were ruled out through further inspection. UVOT also followed up two sources detected by the Zwicky Transient Facility (ZTF): ZTF19aarykkb, ZTF19aarzaod \citep{GCN:24191} and one source from the Panoramic Survey Telescope \& Rapid Response System (Pan-STARRS): AT2019ebq \citep[PS19qp;][]{GCN:24210}. Through further analysis, S190425z was confirmed as BNS with a total mass of $3.4{\rm M_\odot}$ and renamed as GW190425 \citep{abb20b}. 

\subsubsection{S190426c/GW190426\_152155}
Initial classification of S190426c \citep{GCN:24237} gave the probability of this event being: 49 per cent BNS merger, 24 per cent Mass Gap\footnote{Mass Gap implies a system with at least one compact object whose mass is in the hypothetical `mass gap' between NS and BH, taken to be 3 and 5$M_\odot$}, 14 per cent terrestrial and 13 per cent BH-NS merger. However, this classification was updated 4 months later, indicating a 58 per cent probability of being terrestrial in origin \citep{GCN:25549}. Automatic UVOT pipeline analysis of this candidate BNS merger identified 131 Q0 sources. The majority of these were immediately excluded as they were from a couple of images with poor aspect solution which resulted in double or smeared stars. After manual inspection, no UVOT sources were of interest \citep{GCN:24863}. {\it Swift} performed follow-up observations of one external candidate, ZTF19aassfws \citep{GCN:24331}. In GWTC-2, this event is reported as a possible BH-NS merger and renamed GW190426\_152155 \citep{abb20d}.

\subsubsection{S190510g}
This source was initially reported as a high probability BNS merger in \cite{GCN:24442} but was revised more than a day later as having a high likelihood of being terrestrial in origin \citep{GCN:24489}. In that time {\it Swift} uploaded tiled observations of the initial error circle, and identification and inspection of sources had commenced. In those observations, there were 133 Q0 sources. The majority of these were on images with a jump in the aspect resulting in double stars or were false Q0 sources on $uvw1$ images. No sources were considered of interest after manual inspection. However, in \cite{ohg21}, an external candidate was reported: Cand-A09, which was observed by Subaru/Hyper Suprime-Cam. Reviewing the UVOT images it was noted that this source was also observed during the tiling performed by UVOT, however, this source did not pass the pipeline checks in order for a thumbnail to be produced. This GW event, after reanalysis, did not meet the criteria to be included in GWTC-2 \citep{abb20d}. 

\subsubsection{S190718y}
S190718y \citep{GCN:25087} had a high probability of being terrestrial in origin but also had a small chance of being a BNS merger. The {\it Swift} trigger criteria states that any GW events which are flagged as containing a NS would be followed up, so {\it Swift} performed tiling of this GW. Automatic analysis identified 254 Q0 sources and 28 Q1 sources. The vast majority of Q0 sources resulted from two images with jumps in spacecraft position resulting in double images. All but one of these sources were quickly discarded during manual inspection. This GW event, after reanalysis, did not meet the criteria to be included in GWTC-2 \citep{abb20d}.

\subsubsection{S190728q/GW190728\_064510}
S190728q \citep{GCN:25187} had a high chance of being a BBH merger and a 5 per cent chance of one of the remnants having a mass within the mass gap. Automatic analysis identified 2 Q0 sources and 7 Q1 sources, and after manual inspection, only 1 Q0 source of interest remained. In GWTC-2 this event is renamed GW190728\_064510 \citep{abb20d}.

\subsubsection{S190808ae}
S190808ae \citep{GCN:25296} had a high chance of having a terrestrial origin (57 per cent) or being a BNS merger (43 per cent). It was retracted 4.5hr later \citep{GCN:25301}. However, a small number of UVOT tiles were still obtained for this event. Automatic analysis identified 1 Q1 source, which was discarded upon manual inspection.

\subsubsection{S190814bv/GW190814}
S190814bv did not meet the {\it Swift} trigger criteria as it was likely a mass-gap with a large error region \citep{GCN:25324}. The probability region was reduced from 772 deg$^2$ to 38 deg$^2$, within 2 hours after the trigger. Under the O3a trigger criteria for a mass-gap event to be observed, the 90 per cent of the probability in the galaxy-convolved skymap, $P_{0.9}$ must be $\leq10$ deg$^2$ \citep{pag20}. Even with the reduction in the probability region, for S190814bv, $P_{0.9}$ was outside of the trigger criteria with $P_{0.9}=18$ deg$^2$. Therefore a judgement call was made to observe this event. The classification was revised 12 hrs later, resulting in the most likely progenitor being a BH-NS merger \citep{GCN:25333}, instantly meeting the trigger criteria. Automatic analysis identified 15 Q0 sources and 87 Q1 sources. Several of the Q0 sources were spurious detections caused by an image with a jump in the attitude. Upon manual inspection, only 4 Q1 sources were considered to be of interest. UVOT also followed up one source detected by ASKAP: ASKAP 005547-270433  \citep[AT2019osy;][]{GCN:25487}. {\it Swift} observations of S190814bv will also be presented in Cenko et al., (in prep). The LVC confirmed this event as a merger of a BH with a compact object, with it being either the lightest black hole or the heaviest NS yet discovered. This event has been renamed as GW190814 \citep{aab20c}. As this GW event has the most precise localisation of all events observed during O3, extensive searches for an EM counterpart have been performed from the ground too, providing stringent upper limits \citep[e.g.][]{gom19,dob19,and20,ack20}.

\subsubsection{S190822c}
S190822c \citep{GCN:25296} had a high chance of having a terrestrial origin (57 per cent) or being a BNS (43 per cent). It was retracted 4.5hrs later \citep{GCN:25301}. A small number of UVOT tiles were obtained for this event. Automatic analysis identified 1 Q1 source, which was discarded upon manual inspection.

\subsubsection{S190930t}
For S190930t \citep{GCN:25876}, automatic analysis identified 130 Q0 sources and 44 Q1 sources. The vast majority of Q0 sources were due to poor settling of the spacecraft on one tile resulting in double sources. Of the remaining sources, one Q0 source, one Q1 source and one galaxy remained as sources of interest. {\it Swift} also followed-up one source detected by ZTF: ZTF19acbpqlh \citep{GCN:25899}. This single detector GW event, after reanalysis, did not meet the criteria to be included in GWTC-2 \citep{abb20d}.

\subsubsection{S191110af}
S191110af was initially identified as an unmodelled transient candidate \citep{GCN:26222}. However, further investigation of the GW data revealed it to be due to a short period with an elevated rate of instrumental artefacts in the frequency range of the trigger and was retracted \citep{GCN:26250}. Since the retraction was not announced until three days later and the initial GW alert met the {\it Swift} GW trigger criteria, {\it Swift} performed tiled observations of the most probable region. As an unmodelled classification suggests a Galactic origin, {\it Swift} performed 797 tiled observations of the Galactic plane. The number of thumbnails produced was small compared to other GW triggers. A small number of galaxies were serendipitously observed as part of the Galactic plane tiling. The pipeline does not consider the merger classification, therefore, thumbnails were automatically produced for these galaxies and were examined for completeness. Automatic processing found 18 Q0 sources and 24 Q1 sources. Manual inspection of these sources did not reveal any sources of interest.

\subsubsection{S191213g}
For S191213g \citep{GCN:26402}, due to the less than 10 per cent of the probability region being observable within 24 hours no formal tiling was initiated by {\it Swift}, however a few reported sources were followed up. {\it Swift} observed four candidates, three reported by ZTF: ZTF19acykzsk \citep{GCN:26424}, ZTF19acyldun \citep[AT 2019wrt;][]{GCN:26437} and ZTF19acymixu \citep[AT 2019wrr;][]{GCN:26437}, and one reported by Pan-STARRS: PS19hgw \citep[AT2019wxt;][]{GCN:26485}.

\subsubsection{S191216ap}
This event was initially identified as a Mass Gap merger but was subsequently changed to a likely BBH merger \citep{GCN:26454,GCN:26570}. The {\it Swift} GW trigger criteria were not met for this trigger. Therefore no tiling observations were performed. However, a neutrino was reported by IceCube \citep{GCN:26460, GCN:26463}. {\it Swift} performed 100 observations of the overlap between the LVC and IceCube error regions. In addition, a gamma-ray sub-threshold event coincident with LIGO/Virgo and IceCube localisations was reported by HAWC \citep{GCN:26472}. {\it Swift} performed a 7 point tiling, covering 99.6 per cent of the HAWC 0.4 degree radius error circle and a further nine galaxies noted by \cite{GCN:26479} as being consistent with the LIGO-Virgo and HAWC positions \citep{GCN:26498}. In these observations, there were 70 Q0 sources and 4 Q1 sources. However, most of the Q0 sources were due to poor settling on one image resulting in a double image of all objects in the exposure. After manual inspection, only one source was of interest.

\subsubsection{S200114f}
S200114f \citep{GCN:26734} was an unmodelled trigger. The error region was relatively small, and 206 tiles were planned. However, S200115j triggered LIGO-Virgo the following day, which took precedence and prevented the observation plan's completion. Additional analysis of the {\it Swift} follow-up of S200114f will also be presented by Evans et al., (in prep). Automatic analysis of the 69 observed tiles produced thumbnails for 5 Q0 sources and 17 Q1 sources. After manual inspection, one Q0 source remained as a source of interest.

\begin{landscape}
\begin{table}
\centering 
\caption{The main properties of the sources of interest discovered or followed up with the UVOT during the search for the EM counterpart to GWs during the LIGO-Virgo O3 period. The columns are GW alert name; source ID; position in RA and Dec in degrees, J2000; peak UVOT $u$-band magnitude, change in $u$-band brightness, $\Delta m$; Galactic coordinates in degrees, and an initial classification based on a literature search. $\Delta m$ is computed using the peak $u$ mag and either the minimum UVOT $u$-band value, where available or a catalogue $u$ or $g$-band value. Those using a $g$-band archival value are identified with a $g$ in brackets. For the sources discovered with other facilities, we only provide $\Delta m$ when it can be computed either from UVOT observations only or a UVOT $u$-band magnitude and a catalogued $u$-band value exists. For the classifications, we use the term {\it uncatalogued} for sources that have not been observed previously in optical catalogues and {\it unidentified} when an object has archival photometry, but no external classification.
\label{tab:UVOTsrcsmmry}}
\begin{threeparttable}
\begin{tabular}{llccccccc} 
\hline
\hline
GW trigger & Source ID  &     \multicolumn{2}{c}{Position}    & peak (u) magnitude & $\Delta m$ & \multicolumn{2}{c}{Galactic Coordinates} & Classification \\
   	   &     &     RA (deg)  &	Dec (dec)     &       mag (AB)         &      mag &       longitude (deg) &   latitude (deg)      &         \\
\hline
S190412m	   &    Q0\_src10           &	223.91102  &	33.11029      &   $20.32\pm0.28$   &  1.8     & 53.12997  &  62.58951  & candidate AGN \\
S190412m   &	Q0\_src28           &	192.35297  &	14.90385      &   $20.16\pm0.28$   &  1.4     &	300.62138 &  77.76651  & candidate AGN \\
S190412m   &	Q0\_src36           &	214.56987  &	31.35459      &   $20.01\pm0.24$   &  1.0     & 50.71475  &  70.62530  & candidate AGN \\
S190425z   &	Q0\_src136          &	104.61515  &	-45.72216     &   $19.39\pm0.32$   &  $> 0.9$ &	255.76602 &  -17.90450 & uncatalogued \\
S190425z   &	Q0\_src186 	    &	255.58000  &	-12.48562     &   $18.74\pm0.18$   &  $> 2.0$ &	8.37740   &   17.43845 & unidentified  \\	  
S190718y   &	Q1\_src82           &	336.33604  &	-55.99178     &   $20.42\pm0.16$   &  0.7 ($g$) &	334.92714 &  -51.07915 & unidentified  \\
S190814bv  &	Q1\_src5	    &	13.27599   &	-25.38296     &   $20.69\pm0.16$   &  0.9 ($g$) &	135.10517 &  -88.21519 & candidate AGN \\	  
S190814bv  &	Q1\_src49           &	11.34316   &	-24.46617     &   $21.11\pm0.21$   &  1.2 ($g$) &	95.44822  &  -87.00837 & unidentified  \\
S190814bv  &	Q1\_src54           &	13.67442   &	-24.89507     &   $21.33\pm0.17$   &  0.9 ($g$) &	141.26939 &  -87.64981 & unidentified  \\
S190814bv  &	Q1\_src113          &	11.87052   &	-24.22461     &   $20.96\pm0.13$   &  0.6 ($g$) &	105.65106 &  -86.96269 & unidentified  \\	  
S190930t   &	Q1\_src33           &	154.41967  &	34.98805      &   $19.14\pm0.33$   &  $> 1.6$ &	189.78742 &   56.36393 & uncatalogued \\
S190930t   &	Q0\_src93           &	334.96599  &	-48.71116     &   $19.48\pm0.20$   &  2.7     &	346.15130 &  -53.71314 & unidentified  \\
S191216ap  &	Q0\_src147          &	322.0254   &	2.5390	      &   $17.35\pm0.08$   &  4.4     &	55.82826  &  -32.77467 & CV            \\	  
S200114f   &	Q0\_src201          &	109.68957  &	19.74418      &   $17.86\pm0.09$   &  4.9     &	197.97713 &   14.69424 & CV            \\
S200115j   &	Q1\_src1            &	43.28997   &	10.15932      &   $20.30\pm0.28$   &  4.0 ($g$) &	165.43650 &  -42.44842 & unidentified  \\	  
S200115j   &	Q1\_src12           &	36.20419   &	-6.08018      &   $20.62\pm0.20$   &  1.8     &	173.67367 &  -59.41019 & candidate AGN \\	  
S200115j   &	Q1\_src20           &	36.65736   &	-5.86179      &   $20.66\pm0.19$   &  1.0     &	174.01048 &  -58.94003 & candidate AGN \\
S200115j   &	Q1\_src28           &	40.37844   &	-1.37120      &   $20.95\pm0.30$   &  0.9     &	173.26903 &  -53.12805 & candidate AGN \\	  
S200115j   &	Q1\_src39           &	36.19785   &	-7.93239      &   $20.84\pm0.24$   &  6.5     &	176.28980 &  -60.72458 & unidentified  \\	  
S200115j   &	Q1\_src56           &	41.89143   &	5.44519	      &   $21.22\pm0.22$   &  1.3     &	168.03668 &  -47.00212 & unidentified  \\	  
S200115j   &	Q1\_src58           &	44.39037   &	12.77930      &   $21.10\pm0.34$   &  $> 0.1$ &	164.45950 &  -39.71253 & uncatalogued \\	  
S200115j   &	Q1\_src62           &	36.57361   &	-5.49852      &   $21.41\pm0.37$   &  5.6     &	173.40794 &  -58.73583 & unidentified  \\
S200115j   &	Q1\_src78           &	40.32801   &	-2.86688      &   $20.91\pm0.19$   &  1.1     &	174.96031 &  -54.24006 & candidate AGN \\	  
S200115j   &	Q1\_src106          &	36.84194   &	-2.89721      &   $20.79\pm0.27$   &  0.9     &	170.49881 &  -56.63285 & candidate AGN \\ 
S200224ca  &	Q0\_src47           &	176.58794  &	-11.46508     &   $19.64\pm0.23$   &  $> 2.3$ &	278.55108 &   48.30150 & unidentified  \\
S200224ca  &	Q1\_src40           &	173.27717  &	-2.44852      &   $20.44\pm0.34$   &  0.7     &	267.30406 &   54.91160 & candidate AGN \\ 
S200224ca  &	Q1\_src54           &	173.30543  &	-2.46977      &   $20.06\pm0.32$   &  $> 1.4$ &	267.36522 &   54.90807 & uncatalogued \\	  
\hline	   				
S190425z   &	ZTF19aarykkb	   &	258.34145  &     -9.96447      &   $19.43\pm0.18$   &   --  &  12.13348   &   16.55009 &  SN II        \\
S190425z   &	ZTF19aarzaod	   &	262.79149  &     -8.45072      &   $21.82\pm0.30$   &   --  &  15.84702   &   13.59514 &  SN II        \\
S190425z   &	PS19qp		   &	255.32637  &     -7.00289      &   $20.25\pm0.14$   &   --  &  13.08283   &   20.65527 &  SN Ib/IIb    \\
S190510g   &    Cand-A09           &    223.56513  &     4.79320       &   $17.11\pm0.07$   &  $>3.65$ & 1.06508  &   53.19118 &  uncatalogued \\
S190728q   &	ZTF19abjethn       &	326.39538  &	 20.69054      &   $16.41\pm0.06$   &   7.5 &  74.82150   &  -24.35470 &  CV           \\
S190814bv  &    AT2019osy          &     13.94583  &    -27.07583      &   $22.76\pm0.33$   &   --  &  210.07691  &  -89.03152 &  candidate AGN \\
S190930t   &	ZTF19acbpqlh	   &	319.92166  &	 37.52207      &   $20.39\pm0.09$   &	--  &   83.14669  &  -8.492955 &  SN II        \\ 
S191213g   &	ZTF19acykzsk	   &	 32.90454  &     34.04135      &   $19.76\pm0.23$   &   --  &  141.34476  &  -25.94734 &  SN II        \\
S191213g   &	ZTF19acyldun	   &	 79.19999  &     -7.47871      &   $19.40\pm0.14$   &   --  &  208.80094  &  -24.42263 &  SN IIn       \\
S191213g   &	ZTF19acymixu	   &	 90.91394  &     60.72825      &   $> 21.09   $     &   --  &  153.11430  &   17.92000 &  SN Ia        \\
S191213g   &	PS19hgw		   &     28.92475  &     31.41789      &   $19.14\pm0.20$   &   --  &  138.67292  &  -29.48205 &  SN IIb       \\
S200213t   &	ZTF20aamvmzj	   &	 27.18921  &     51.43048      &   $20.08\pm0.27$   &   $>0.97$  &  131.95909  &  -10.43382 &  uncatalogued \\
S200213t   &    ZTF20aanakcd       &      8.15705  &     41.31573      &   --$^*$             &   --  &  119.13951  &  -21.41749 &  SN IIn       \\
\hline
\end{tabular} 
  \begin{tablenotes}
    \item[* No observations were taken in the UVOT $u$-band] 
  \end{tablenotes}
\end{threeparttable}
\end{table}
\end{landscape}

\subsubsection{S200115j}
S200115j \citep{GCN:26759} was classified as a Mass Gap merger. The localisation skymap changed considerably between the initial BAYESTAR and later LALInference maps, with the error region shifting and decreasing in size; {\it Swift} observations were planned and initiated when only the BAYESTAR maps were available. Automatic analysis resulted in 3266 thumbnails from 512 UVOT images, with 15 Q0 sources and 110 Q1 sources. In manual inspection, 10 Q1 sources remained of interest, with additional follow-up observations performed.

\subsubsection{S200213t}
S200213t was identified as a BNS merger (63 per cent) with a non-negligible chance of being terrestrial in origin \citep[37 per cent; ][]{GCN:27042}. The {\it Swift} GW trigger criteria were not met for this trigger. Therefore no tiling observations were performed. However, IceCube announced the detection of 1 neutrino \citep{GCN:27043} that was temporally and spatially coincident with the GW. {\it Swift} performed seven tiling observations within the error region of this neutrino event. Within these tiles, 3 Q0 sources were automatically identified but were rejected upon manual inspection. {\it Swift} also followed up two ZTF sources: ZTF20aamvmzj \citep[AT2020cja;][]{GCN:27051} and ZTF20aanakcd \citep[AT2020cmr;][]{GCN:27068}.

\begin{table*} 
\centering 
\caption{UVOT observations. A short summary is provided here. The full table is given in the Supplementary Table S.1.
\label{tab:UVOTobs}}
\begin{tabular}{llcccc} 
\hline 
GW trigger & Source of Interest   & Start Time (UT)           & Filter & Exposure (s) & Magnitude (AB) \\
S190412m	&  Q0\_src10	&   2019-04-12T19:31:45	&  $u$   	&  79	& $20.32\pm0.28$ \\ 
S190412m	&  Q0\_src10	&   2020-06-26T12:03:44	&  $u$   	&  543	& $20.47\pm0.12$ \\ 
S190412m	&  Q0\_src28	&   2019-04-12T17:48:36	&  $u$   	&  75	& $20.16\pm0.28$ \\ 
S190412m	&  Q0\_src28	&   2020-06-30T08:38:30	&  $u$   	&  438	& $20.40\pm0.15$ \\ 
S190412m	&  Q0\_src36	&   2019-04-12T21:33:49	&  $u$   	&  77	& $20.01\pm0.24$ \\ 
S190412m	&  Q0\_src36	&   2020-06-27T00:42:09	&  $u$   	&  518	& $20.32\pm0.12$ \\ 
\hline
\hline 
\hline 
\end{tabular} 
\end{table*} 

\begin{table}
\centering 
\caption{The main X-ray upper limits measured with {\it Swift}/XRT of the sources of interest discovered or followed-up with the UVOT during the search for the EM counterpart to GWs during the LIGO-Virgo O3 period. The columns are GW event; Source ID; XRT count rate upper limit in 0.3-10 keV energy range. The count to observed (absorbed) flux conversion factor is $\sim 4.3\times10^{-11}\,{\rm erg\,cm^{-2}\,count^{-1}}$, assuming a power-law index of $\Gamma=1.7$ and an absorbing column of $N_H = 3\times10^{20}\,{\rm cm}^{-2}$.}
\label{tab:XRTsrcsmmry}
\begin{tabular}{llcc} 
\hline
\hline
GW trigger & Source ID   & XRT Upper Limits  \\
   	   & 	 	        & 0.3-10keV            \\
   	   &                        &     (c/s)            \\
 \hline
S190412m   &	Q0\_src10		&		5.97E-02       \\
S190412m   &	Q0\_src28		&		1.63E-01       \\
S190412m   &	Q0\_src36		&		1.10E-01       \\
S190425z   &	Q0\_src136		&		1.79E-01       \\
S190425z   &	Q0\_src186		&		8.43E-03       \\
S190718y   &	Q1\_src82		&		1.48E-02       \\
S190814bv  &	Q1\_src5		&		2.49E-02       \\
S190814bv  &	Q1\_src49		&		1.21E-02       \\
S190814bv  &	Q1\_src54		&		1.54E-02       \\
S190814bv  &	Q1\_src113		&		1.05E-02       \\
S190930t   &	Q1\_src33		&		1.13E-01       \\
S190930t   &	Q0\_src93		&		2.64E-04       \\
S191216ap  &	Q0\_src147		&		8.01E-02       \\
S200114f   &	Q0\_src201		&		1.08E-01       \\
S200115j   &	Q1\_src1 		&		1.95E-02       \\
S200115j   &	Q1\_src12 		&		9.47E-03       \\
S200115j   &	Q1\_src20		&		1.35E-02       \\
S200115j   &	Q1\_src28		&		5.83E-03       \\
S200115j   &	Q1\_src39		&		3.90E-03       \\
S200115j   &	Q1\_src56		&		1.13E-02       \\
S200115j   &	Q1\_src58		&		5.29E-03       \\
S200115j   &	Q1\_src62		&		4.31E-03       \\
S200115j   &	Q1\_src78		&		7.45E-03       \\
S200115j   &	Q1\_src106		&		4.52E-03       \\
S200224ca  &	Q0\_src47		&		4.75E-03       \\
S200224ca  &	Q1\_src40		&		1.04E-02       \\
S200224ca  &	Q1\_src54		&		7.90E-03       \\
\hline
\hline
S190425z   &	ZTF19aarykkb		&		2.50E-03       \\	
S190425z   &	ZTF19aarzaod		&		3.67E-03       \\	
S190425z   &	PS19qp			    &		1.16E-02       \\
S190510g   &    Cand-A09            &       1.30E-01       \\
S190728q   &	ZTF19abjethn       	&		--	           \\
S190814bv  &    AT2019osy          	&		1.96E-03       \\
S190930t   &    ZTF19acbpqlh       	&		5.28E-03       \\
S191213g   &	ZTF19acykzsk		&		6.47E-03       \\
S191213g   &	ZTF19acyldun		&		4.65E-03       \\
S191213g   &	ZTF19acymixu		&		5.04E-03       \\
S191213g   &	PS19hgw			    &		1.89E-03       \\
S200213t   &	ZTF20aamvmzj		&		6.71E-04       \\
\hline
\hline
GW170817   &    AT2017gfo/GRB170817A &       2.8E-04     \\
\hline
\hline
\end{tabular} 
\end{table}

\subsubsection{S200224ca}
S200224ca \citep{GCN:27184} was identified as a BBH merger ($>99$ per cent). Due to the relatively good localisation, with 50 (90) per cent area within 13 (72) deg$^2$, this GW met the {\it Swift} trigger criteria. Automatic analysis found 17 Q0, 65 Q1 sources. Of these, 1 Q0 source and 2 Q1 sources were manually identified as sources of interest. A discussion of {\it Swift} observations of S200224ca is also presented in \cite{kli20}.

\subsubsection{S200225q}
S200225q \citep{GCN:27193} was classified as a BBH merger (96 per cent). S200225q did not formally meet the $\it Swift$ trigger criterion. However, the LALInference skymap released 38 hr later is well localised, so it was decided to do a 37-point tile to cover the 50 per cent region. This plan was interrupted by GRB 200227A \citep{GCN:27234} and a second plan was initiated. A total of 70 tiles performed across the two plans. The second phase of 500 s tiles was not performed due to the delay. No Q0 or Q1 sources were identified by the UVOT pipeline.

\subsection{Examination of the sources of interest}
Altogether {\it Swift}/UVOT found or followed up 40 sources of interest, 36 of which have more than one detection, either from archives or from follow-up with UVOT or other facilities at the time of detection. The main properties of these sources are listed in Table \ref{tab:UVOTsrcsmmry} and their photometry is provided in Table \ref{tab:UVOTobs}. In the following, we investigate the properties of these sources. In this analysis, we also include AT2017gfo, the EM counterpart of GW170817, for comparison. Using catalogue information reported by VizieR, we were able to divide these 40 sources into five initial classifications: candidate AGN - source reported in the literature as confirmed or candidate AGNs/quasars (QSOs); SNe - sources that are reported as or confirmed as a supernova; CV - sources identified as Cataclysmic variables; unidentified - sources that have archival photometry, but are not identified; uncatalogued - sources that have no archival photometry and have no identification. Dividing the sources into these classes, we have 11 candidate AGN/QSOs, 3 CVs, 9 SNe, 11 unidentified sources and 6 uncatalogued sources. 

\subsubsection{Positions of the sources of interest}
\begin{figure}
\includegraphics[angle=0,scale=0.41,trim={2.2cm 3.5cm 2.cm 8.4cm},clip]{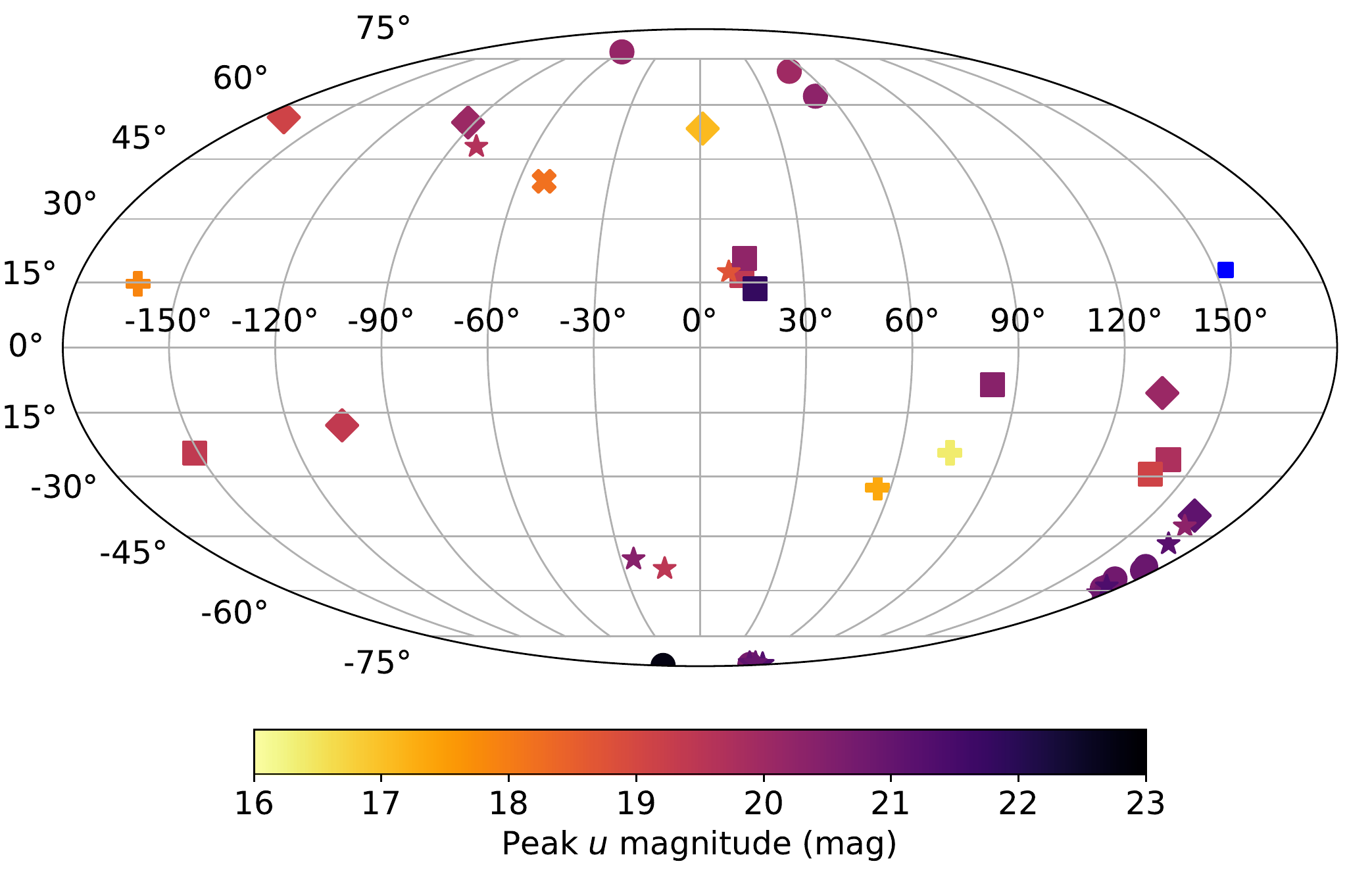}
\includegraphics[angle=0,scale=0.41,trim={2.2cm 2.0cm 2.cm 8.4cm},clip]{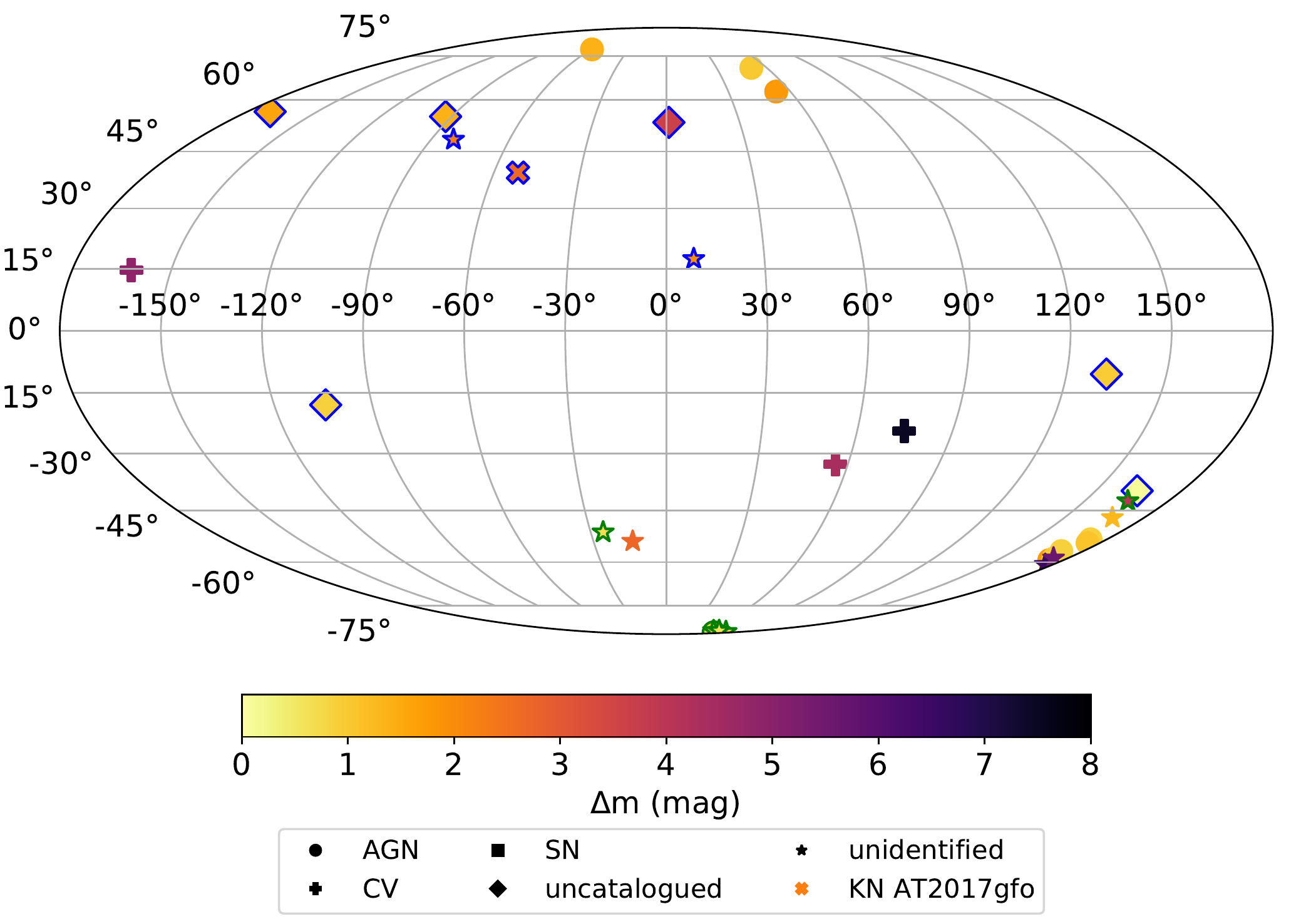}
\caption{The positions on the sky in galactic coordinates of the 40 sources of interest detected or followed up by UVOT during O3. The colours represent the peak $u$-band magnitude (AB; top panel) and the change in magnitude, $\Delta m$ (bottom panel), where $\Delta m$ is calculated using two UVOT $u$ exposures or the UVOT detection magnitude and an archival $g$-band magnitude. For the panels displaying peak $u$ magnitudes, we have coloured objects blue where UVOT measured only an upper limit. For the panels displaying $\Delta m$, we use a blue edge to denote lower limits and a green edge to denote where change was calculated using UVOT $u$ and an archival $g$-band magnitude. Different symbols separate the sources into five categories: candidate AGN, supernovae, cataclysmic variables, unidentified and uncatalogued with the key given in the legend.}
\label{src_positions}
\end{figure}

In Fig. \ref{src_positions} we display the positions of the 40 sources of interest on the sky in galactic coordinates. In the first panel, we colour the data points by peak $u$-band luminosity and in the second panel by the apparent change in magnitude ($\Delta m$). The sources are also divided by their initial classifications as given in the legend. There is no apparent correlation between source position and peak $u$-band magnitude or source position and $\Delta m$. Only one source of interest, a SN, is within $\pm 10\deg$ of the Galactic plane. 

\subsubsection{Temporal behaviour of the sources of interest}
\label{temp_beh}
In Supplementary Fig. S.1-S.7 we display the light curves of the sources of interest that have more than one photometric detection from UVOT and other facilities and archives. We have not produced light curves for those sources initially classified as SNe. We display the light curves so that sources from the same class are grouped. We have collated optical photometry available in the Sloan Digital Sky Survey \citep[SDSS;][]{ala15}, Canada-France-Hawaii Telescope Legacy Survey \citep[CFHTLS;][]{cui12}, Dark Energy Survey \citep[DES;][]{abb18}, Pan-STARRS \citep[][]{cha16}, Palomar Transient Factory \citep[PTF;][]{law09}, ZTF \citep[][]{bel19,gra19} and Catalina Real Time Transient Survey \citep[CRTS;][]{dra09}\footnote{We cross-matched using a search radius of $2\arcsec$ for all optical catalogues except for CRTS which we used $3\arcsec$.}. 

Light curves could be produced for 10 of the 11 candidate AGN; a light curve for AT2019osy could not be created. This class has the light curves with the most photometric data points and includes light curves that span several years. Small changes in $\Delta m$ occur on short time scales, and larger changes occur over several years. These ten light curves can be approximately divided into those that vary around a fixed magnitude with only small scale variability ($\Delta m \sim 0.5 $mag), and those for which tend to have this low level, short time scale variability, but also larger, longer-term variability ($\Delta m\sim 1-2$ mag). For the CVs, we were able to produce light curves for 2 of the 3. The data are sparse but show large changes in magnitude that occur on timescales shorter than the AGN. For the most part, the light curves are poorly sampled for the unidentified sources compared to those of the candidate AGN, although the light curves span similar $>1000$ day timescales. The typical magnitude range of the unidentified sources, $\sim 19-22$, is similar to that observed for the candidate AGN. The light curves of both Q1\_src39 (S200115j) and Q1\_src62 (S200115j) are of particular interest since the minimum archival values from CFHTLS are $\sim 27$ mag in $u$-band, while the UVOT detection is $\sim21$ mag. There are also archival detections by other facilities at those locations suggesting that either these sources have slowly brightened over several years or that they intermittently brighten by several magnitudes at a time. A similar large magnitude change is also observed for Q1\_src1 (S200115j), with archival $g$-band magnitude of 24.27 mag. For this event, a rapid two magnitude change in the $i$-band was caught by Pan-STARRS on a very short timescale, around 20min. For the uncatalogued sources, none of those first detected by UVOT have enough photometry available to produce light curves. We were, however, able to produce a light curve for one uncatalogued source, which ZTF initially identified during S200213t: ZTF20aamvmzj, see Supplementary Fig. S.7. Interestingly, the $r$ and $i$ data reported by \cite{kas20} is approximately flat between the two epochs, whereas the UVOT data appears to decay. By normalising all filters to the $v$-band, we measure a decay slope of $-1.45\pm0.23$. This is in sharp contrast to the decay slope of $\alpha = 0.04$ in the $r$-band reported by \cite{kas20}. This difference suggests a fast decay of the source at UV wavelengths.

\begin{figure}
\includegraphics[angle=0,scale=0.65,trim={0.7cm 0.cm 0.5cm 0cm},clip]{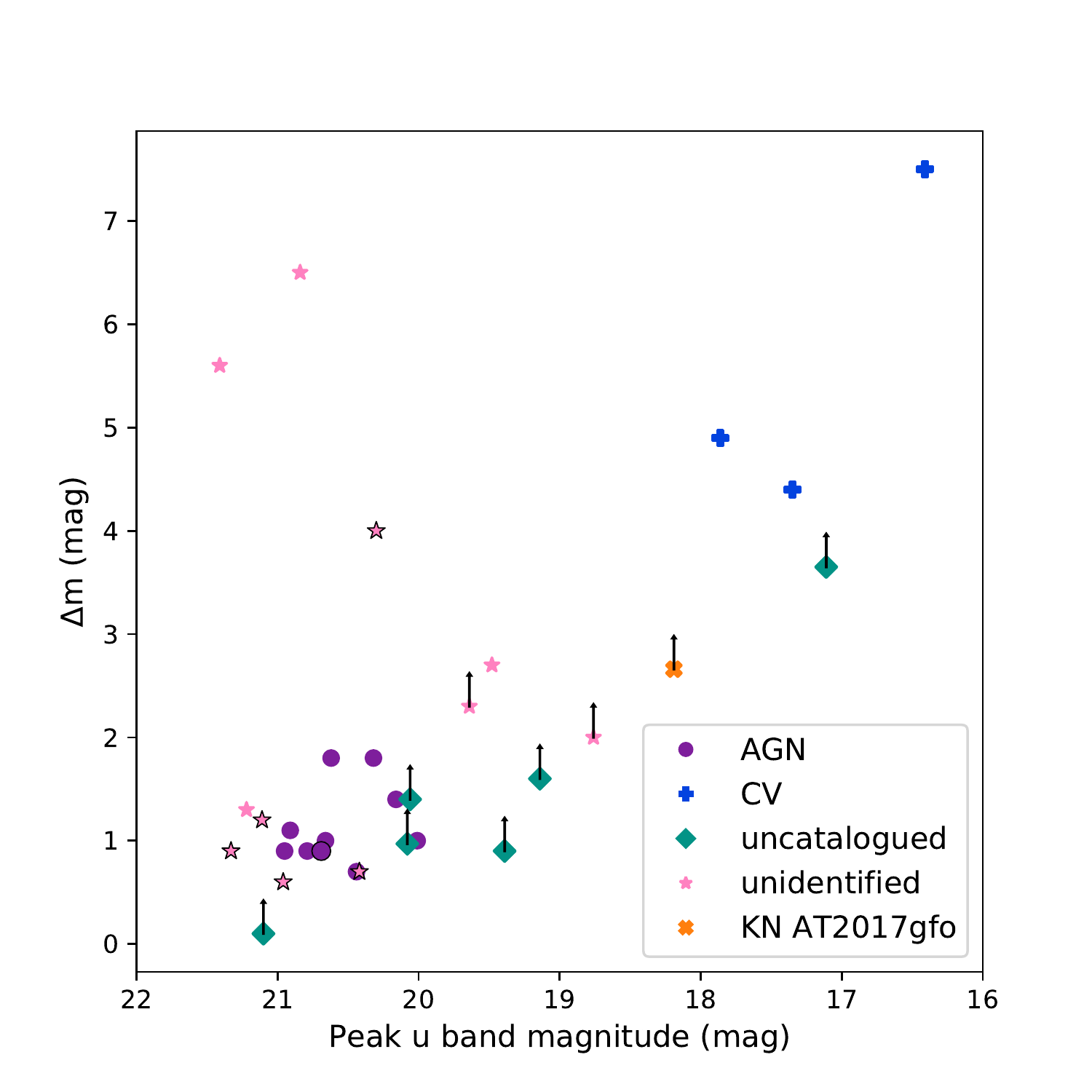}
\caption{The peak $u$-band magnitude (AB) versus the change in magnitude, $\Delta m$. $\Delta m$ is calculated using two UVOT $u$ exposures or the UVOT detection magnitude and an archival $g$-band magnitude. The sources displayed are divided into the categories: candidate AGN, cataclysmic variables, unidentified and uncatalogued, see legend for colours and symbols. Arrows indicate lower limits to $\Delta m$, and points with a black outline indicate those point where $\Delta m$ was calculated using an archival $g$-band magnitude rather than a UVOT $u$ band image. We also display the peak $u$-band magnitude vs $\Delta m$ for AT2017gfo, the EM counterpart to GW170817.}
\label{peak_delta}
\end{figure}

In Fig. \ref{peak_delta} we display the peak $u$-band magnitude against $\Delta m$. Those sources identified as candidate AGN are clustered between 20th and 21st magnitude with $\Delta m$ between 0.5 and 2 magnitudes. The three CVs lie in the top right corner having the brightest peak $u$-band magnitudes and the largest $\Delta m$. The sources classified as unidentified appear to cluster into three areas: one group are similar to that of the AGN, but on average are fainter ({$\ge$}21 mag), the second group lie to the top right of the AGN, they are brighter between 19th and 20th mag and have $\Delta m >$ 1.5 mag (Q0\_src186, Q0\_src93 and Q0\_src47; from GW events S190425z, S190930t, S200224ca, respectively), which lie at a similar position at AT2017gfo, and the third group lies in the top left corner, consisting of three sources with $\Delta m >$ 4 and low peak $u$-band values (Q1\_src1, Q1\_src39 and Q1\_src62 all from S200115j). Five of the uncatalogued sources have $\Delta m <$ 2 and are between 19 and 21.5 mag. The sixth uncatalogued source appears to be as bright and have as large a change in magnitude as the CVs. No SNe are displayed since no $\Delta m$ can be computed from the available data.

\begin{figure}
\includegraphics[angle=0,scale=0.65,trim={0.5cm 0.cm 0.5cm 0cm},clip]{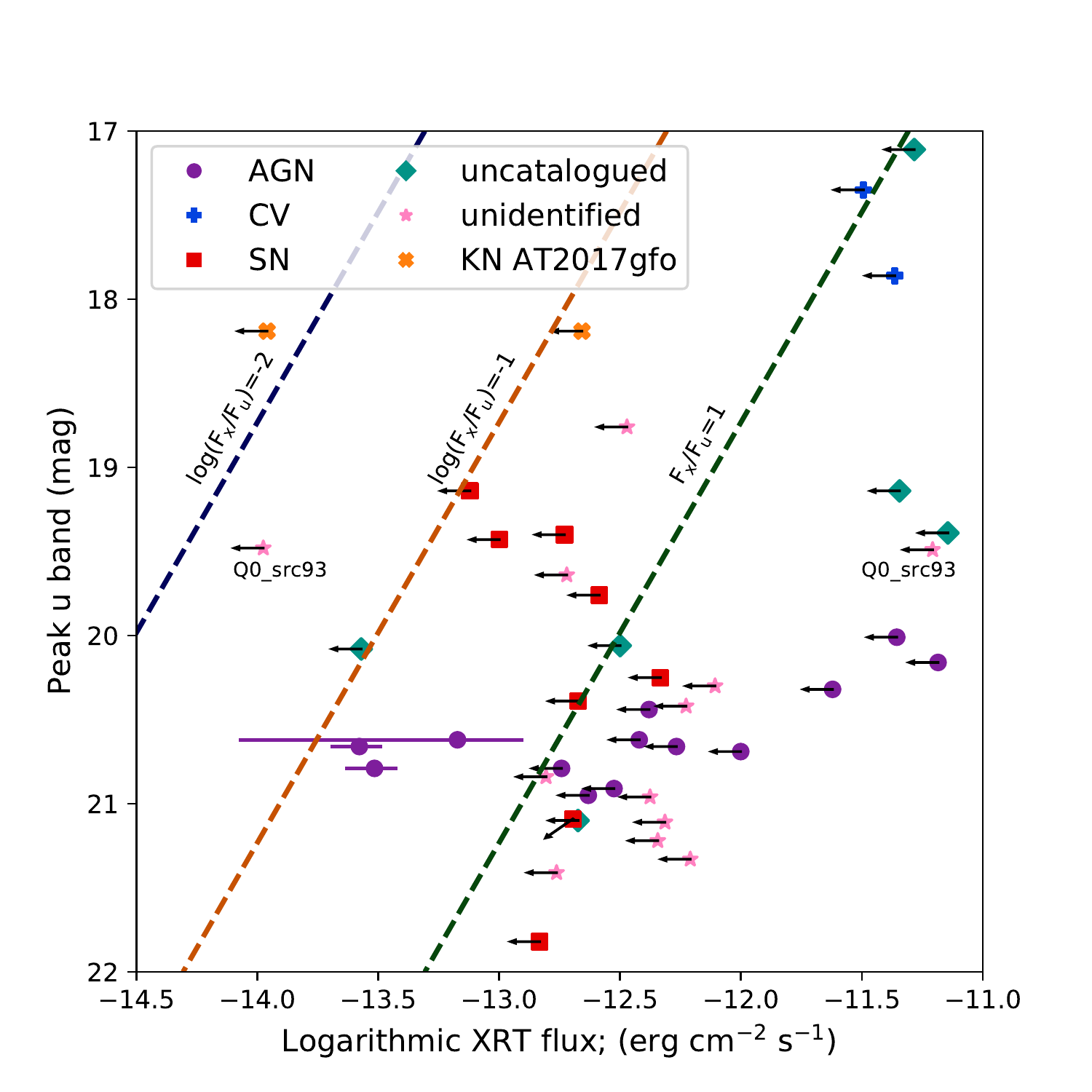}
\caption{The X-ray flux (erg cm$^{-2}$ s$^{-1}$) versus the peak $u$-band magnitude. The X-ray flux values measured by XRT are all in the energy range 0.3-10\,keV and are all 3$\sigma$ upper limits. These values were determined from the XRT count rate using a count to observed (absorbed) flux conversion factor of $\sim 4.3\times10^{-11}\,{\rm erg\,cm^{-2}\, count^{-1}}$, calculated assuming a power-law index of $\Gamma=1.7$ and an absorbing column of $N_H=3\times10^{20}\,{\rm cm}^{-2}$. For AT2017gfo associated with GW170817 and for Q0\_src93, we provide two points with two different XRT upper limits: one upper limit derived from the first observation and the second deeper limit is determined from a stack of all XRT observations. In this figure, we also include archival {\it XMM-Newton} (0.2-12 keV; purple) flux values for three of the candidate AGN sources \citep{web20}. The corresponding XRT upper limits for these three sources are shallower and are immediately to the right. The sources are separated into five initial categories: candidate AGN (circle), SNe (square), CVs (plus), unidentified (star) and uncatalogued (diamond). Sources with diagonal arrows indicate that both the X-ray flux and the peak $u$-band magnitude are upper limits. The diagonal dotted lines represent lines of proportionality with the optical flux 0.1 (blue), 10 (orange) and 100 (green) times the X-ray flux, with the ratio $F_u/F_X$ computed using the optical flux in $\nu F\nu$ and the X-ray flux in the 0.3-10\,keV energy range.}
\label{peak_x-ray}
\end{figure}

In Fig. \ref{peak_x-ray}, we display the peak $u$-band magnitude against the observed XRT X-ray (0.3-10 keV) flux, where the observed X-ray flux is computed using a power-law index of $\Gamma=1.7$ and an absorbing column of $N_H = 3\times10^{20}\,{\rm cm}^{-2}$. Three of the candidate AGN have archival X-ray detections from {\it XMM-Newton} \citep{web20}, which we also include. Since none of the sources are detected by XRT, all the XRT flux values are upper limits. For the most part, these upper limits are reasonably shallow. Only two sources lie below $log (F_x/F_u) =-1$, one uncatalogued source (ZTF20aamvmzj; S200213t) and one unidentified source (Q0\_src93; S190930t), likely excluding these two sources as resulting from an AGN. AT2017gfo, the EM counterpart of GW170817, has an even sharper contrast between the X-ray and optical flux with $log (F_x/F_u) <-2$.

\begin{figure}
\includegraphics[angle=0,scale=0.6, trim={0.0cm 0.5cm 0cm 0.4cm},clip]{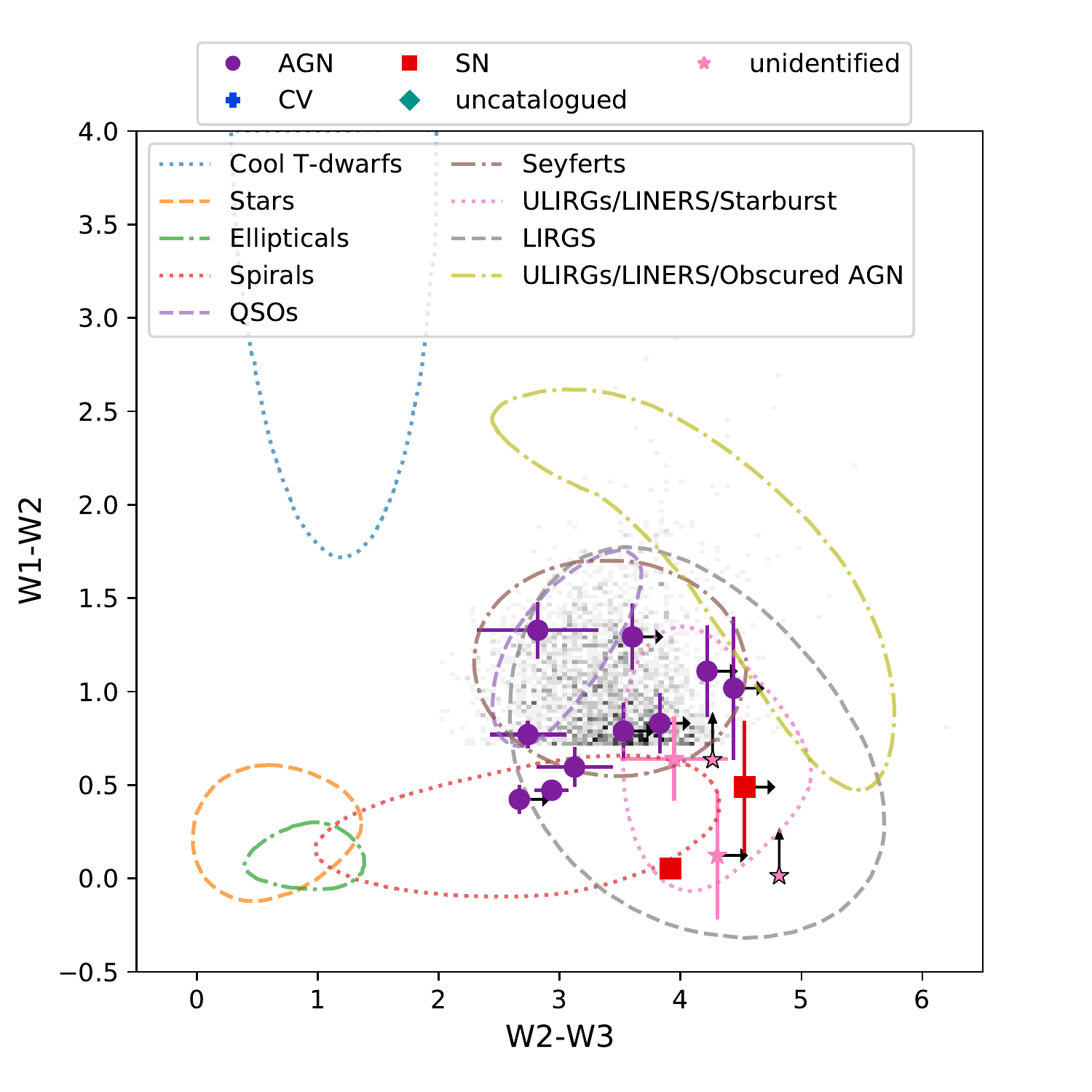}
\caption{Colour-colour diagram of ALLWISE catalogued sources cross-matched with the UVOT sources of interest. We show the UVOT sources of interest separated into five initial categories (see legend) along with a histogram of $10^5$ of the 35~$\times$~10$^6$ ALLWISE AGN candidates from \protect\cite{ass18}, selected using a completeness criteria of 75 per cent (grey scale). In addition, we have overlaid the expected locations of different classes of WISE objects as given in Fig 12 of \protect\cite{wri10}. The two points outlined in black are detected in the WISE W1 filter only. Most of the UVOT sources of interest show consistency with the LIRG class of objects.} 
\label{ALLWISE_W1W2_W2W2}
\end{figure}

We have also cross-matched our sources against the ALLWISE catalogue \citep{cut14} using a search radius of $6\arcsec$. In Fig. \ref{ALLWISE_W1W2_W2W2} we display W2-W3 versus W1-W2 colours of 16 sources of interest. Overlaid on this figure are regions indicating locations of particular types of IR objects \citep{wri10}, we also display 10k AGN determined using the completeness criteria of 75 per cent from \cite{ass18} that also have detections in the three WISE bands. The sources of interest are inconsistent with being stars, cool T-dwarfs, ellipticals and obscured AGN. Almost all of the sources for which there is a WISE match are found in the luminous infrared galaxy (LIRG) colour region. The sources in this region are a mixture of AGN (Seyferts and QSOs) and star-forming galaxies. This is strong evidence that all of the transient sources with WISE matches are extragalactic. Some of these transient sources are candidate AGN, while others are SNe, with the WISE colours indicating that they are located in star-forming galaxies. The unidentified sources with WISE matches are likely to fall into one of these two categories.

\begin{figure*}
\includegraphics[angle=0,scale=0.5, trim={0cm 0.cm 1.5cm 0.cm},clip]{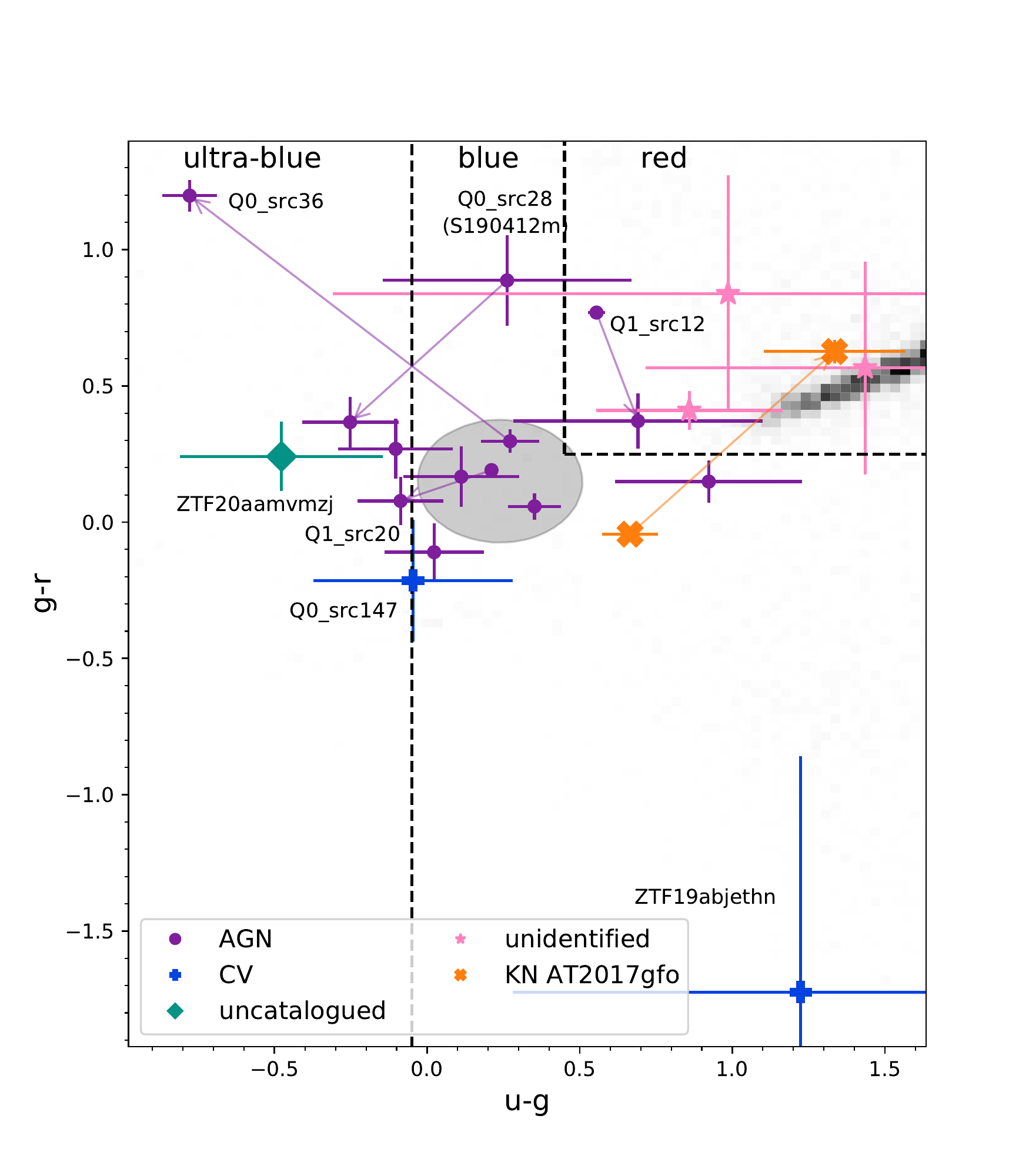}
\includegraphics[angle=0,scale=0.5, trim={0cm 0.cm 1.5cm 0.cm},clip]{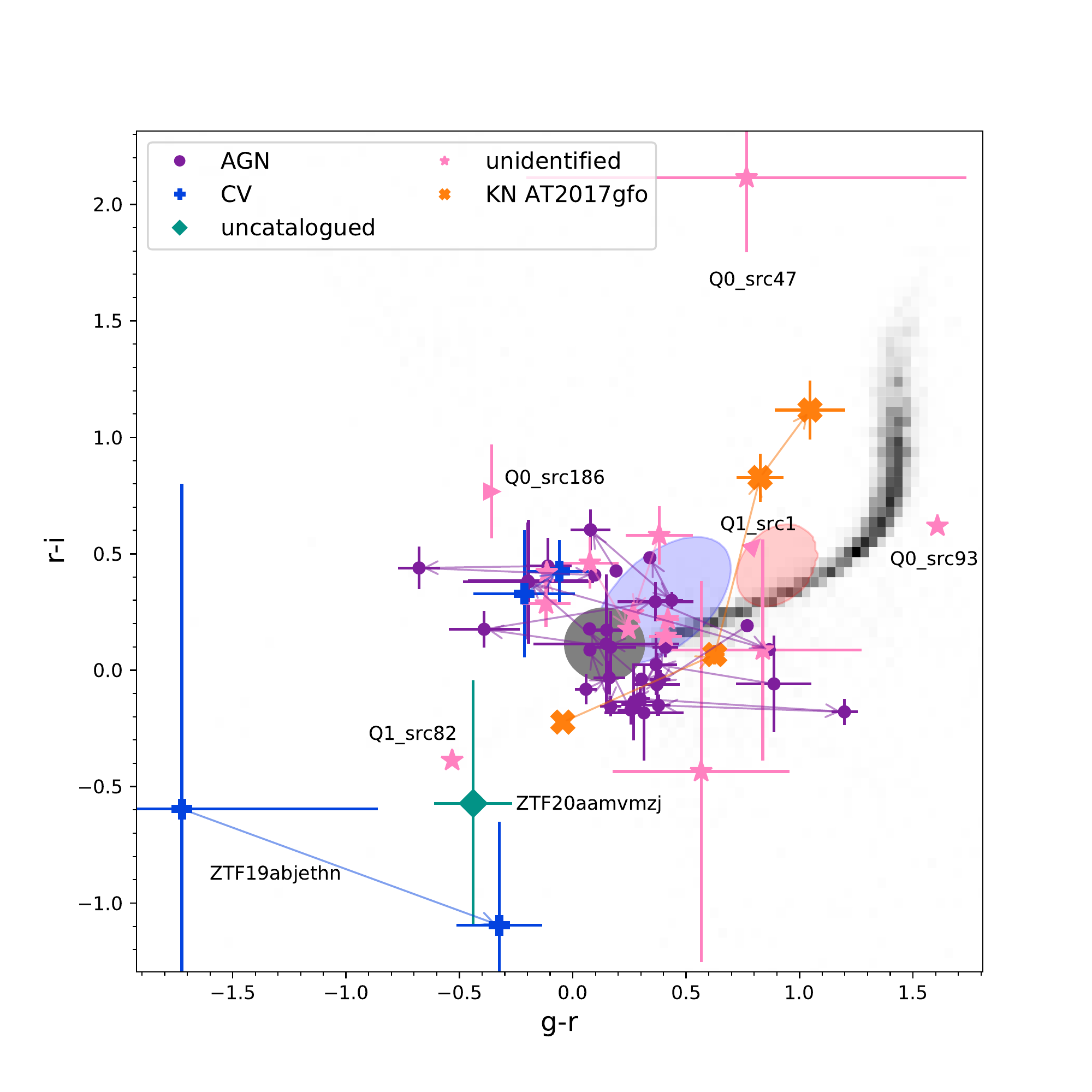}
\caption{Colour-colour diagram of the archival photometry of the sources of interest, separated into the five initial categories (see legend), with photometry taken from the SDSS, CFHTLS, the Pan-STARRS stacked exposures and DES catalogues. Left: $u-g$ versus $g-r$. Right: $g-r$ versus $r-i$. In the left panel, the dotted lines divide the figure into regions identifying objects as red, blue and ultra-blue, adapted from \protect\cite{law16}. In both panels, the grey region indicates the colour location of 90 per cent of SDSS spectroscopic quasars \protect\citep[see][for details]{law16}, and we display a 2-D histogram, given in grey, of the SDSS colours of 10,000 stars from the GAIA DR2 catalogue, selected from a region at high Galactic latitude. In the right panel, the blue region represents the location of the blue cloud galaxies, and the red region represents the red sequence galaxies, both out to $z = 0.22$ \protect\citep[adapted from][]{law16}. In each panel, arrows connect points of the same source in chronological order. In the left panel, these are labelled. In both panels, we have labelled sources that are of particular interest as they are outliers compared to the majority of points. In both panels, we identify AT2017gfo, the EM counterpart to GW170817, in orange. In the right panel, two sources have one filter with no detection, one in the $g$-band and the other in the $r$-band. These points have triangles pointing away from their limit. These colours have been corrected for Galactic extinction. We have not corrected for host extinction for those sources external to our Galaxy. Correction for host extinction would move points down and to the left in both panels. }
\label{Optical_colours}
\end{figure*}

In Fig. \ref{Optical_colours} we provide two plots displaying the colours of the objects. We have used archival photometry from SDSS, CFHTLS, DES and stacked exposures from Pan-STARRS. The first panel, adapted from \cite{law16}, displays $u-g$ versus $g-r$ with regions identifying objects as red, blue and ultra-blue. The grey region indicates the colour location of 90 per cent of SDSS spectroscopic quasars \citep[see][for details]{law16}. In addition, we also display a 2-D histogram, given in grey, of the SDSS colours of 10,000 stars from the GAIA DR2 catalogue, selected from a region at high Galactic latitude. Of the sources of interest, six are found in one or more catalogues, and these points are joined by red lines, indicating spectral evolution and that colour depends on the time of observation. We label some of the sources that may be of interest to the reader. Some of the candidate AGN are consistent with the quasar region, while the others are more scattered in their position in colour. Of the candidate AGN, two are of particular interest, both of which have two data points. Q0\_src36 (S190412m) moves between being blue and ultra-blue, while Q1\_src12 (S200115j) is the only AGN consistent with being red. Four other sources lie in the red section. Three of these are Q1\_src39, Q1\_src56, Q1\_src62 (all from S200115j); however, the errors on these are large. The fourth object is AT2017gfo, the KN associated with GW170817, which is initially blue and later becomes red. The second panel displays $g-r$ versus $r-g$. In this panel, we also display a grey region typical of the colours of quasars and the grey histogram providing the colours of stars. The blue region represents the location of the blue cloud galaxies, and the red region represents the red sequence galaxies, both out to $z = 0.22$ \citep[adapted from][]{law16}. Due to more facilities observing in $g$, $r$ and $i$, there are more sources of interest with archival observations and many sources are found in more than one catalogue; red lines connect these. Most of the points avoid the red region, which corresponds to low-redshift elliptical galaxies. This suggests that apart from the AGN (for which we can say little about the host galaxies), the extragalactic transients identified by our survey are primarily located in star-forming galaxies. This observation may prove useful in our follow up to future GW observing runs. The UVOT $u$-band survey is almost devoid of serendipitous transients in nearby elliptical galaxies, but we expect that some BNS mergers take place in these environments, because short GRBs have been observed to come from elliptical galaxies as well as star-forming galaxies \citep[e.g.][]{geh05,fon13}. Therefore UVOT transients found in elliptical galaxies should be prioritised in future GW observing runs.

\section{Discussion}
\label{discussion}
Overall, 18 LIGO/Virgo GW triggers were followed up at differing extents by {\it Swift}/UVOT. Of these 18 candidate GW events, four were initially classified as BBH triggers, six as BNS events, two each of BH-NS and Mass Gap triggers, one unmodelled/burst trigger, and the remaining three were subsequently retracted. Of the ten events occurring in O3a, only 5 meet the criteria to be included in the GWTC-2 catalogue of O3a \citep{abb20d}. Within the 6.4k UVOT tiles obtained by {\it Swift} across the 18 GW events, automatic analysis found a total of 828 Q0 sources, 344 Q1 sources, and after manual inspection, 9 Q0 and 18 Q1 sources were considered to be of interest. None of these were considered to be EM counterparts of the GW events when they were detected. In the following, we will discuss the importance of UVOT in the search for the EM counterpart to GWs, explore the serendipitously discovered sources of interest and discuss future improvements that can be made to UVOTs performance in detecting and following EM counterparts to GWs. 

\subsection{Expected optical/UV emission from KNe and UVOT}
Detecting optical/UV (blue) emission from future KNe can be a particularly useful tool in determining the properties of the merged binary system. It is expected that ejecta with electron fraction $Y_e>0.25$ will produce the blue part of the KN. The merger remnant and its duration of survival are linked to the total binary mass of the merger \citep{met19}. The longer the merger remnant survives, the larger the expected ejecta mass with electron fraction $Y_e>0.25$ and thus the brighter the KN in the optical/UV \citep{met14,per14,lip17,met17}. For BNS mergers with the largest total binary mass, the remnant will undergo prompt collapse to a BH. The fraction of material with $Y_e>0.25$ is expected to be small, and the resulting KN is expected to be red and dimmer than AT2017gfo. For increasingly smaller total BNS masses, the merger remnants are expected to be hypermassive (HMNS), supermassive (SMNS) remnants \citep{shi19,met19}, for which the stability before collapse is between 30s to hundreds of milliseconds; and stable NSs, which resist collapse completely \citep{kas15,met19,mar19}. It is expected that the most numerous remnants will be HMNS \citep{mar19,met19}, which are expected to produce blue emission. For BH-NS, the NS is expected to be swallowed whole, resulting in no/little observable emission \citep{shi19} or if the NS is disrupted, the emission expected to be predominantly red, but possibly with some blue emission \citep{jus15,met19}. BH-NS mergers are expected to have higher dynamical ejecta mass compared to BNS mergers \citep{fer16,met19} and thus, the light curves are expected to be more luminous. However, on average BH-NS mergers are expected to be detected out to larger distances than BNS mergers \citep{abb20f}, and so the benefit to an observer of the increase in luminosity may be more than offset by the larger expected source distance \citep{met19}.

The blue emission can also provide more specific diagnostics, providing information on the geometry of the binary system, the emission mechanism and the ejecta composition. For instance, a blue KN may only be present for a fraction of KNe; whereby the viewing angle is optimised \citep[cf.][]{chr19}: for viewers close to the equatorial plane, the blue emission may be blocked by the high-opacity lanthanide-rich tidal ejecta \citep{kas15}, whereas close to the binary rotation axis both blue and red parts of the KN may be observed. Several emission mechanisms can produce optical/UV emission in KNe, such as a form of wind produced by the accretion disk, neutrino driven or from the magnetised NS \citep{met18}, or the blue emission may be produced dynamically by the ejecta, by the tidal tail or through shock heating of the ejecta. NS remnants with strong magnetic fields may also produce optical/UV emission that may be brighter than a radioactively powered KN \citep{met14b,met19,fon21}.
Therefore additional observations of KNe in the blue are essential to understand the range of sources that can produce this emission \citep{met19}. The early blue emission within the first hours may also be enhanced by heating caused by free neutrons in the outer mass layer of the ejecta \citep{met15}. It is therefore essential to obtain early, well-sampled multi-wavelength observations of KNe in order to distinguish between the variations in behaviour due to differences in the merging system, properties of the in-going binary, and viewing angle relative to the binary inclination \citep{met19}. 

\subsubsection{Impact of UVOT on EM counterpart Searches}
UVOT has many advantages over ground-based instrumentation in detecting and observing the optical/UV emission from KNe. {\it Swift} is the only satellite capable of performing a fast response targeted tiling campaign. Under optimal conditions, {\it Swift} may be re-pointed to begin tiled observations of the GW error region within an hour of detection. UVOT can observe 24 hours a day and is not limited by the weather or a restricted viewing angle beyond that imposed by the Earth and Sun. {\it Swift}/UVOT is, therefore, able to search for the EM counterpart to GW alerts in portions of the sky, which are restricted or of limited visibility to ground-based observers. Compared to the other facilities observing during O3 \citep[e.g.][]{and19,gol19,lun19,ack20,and20,ant20,gom20,kas20,pat20,tha20,vie20,ana21,ohg21}, {\it Swift}/UVOT was consistently one of the fastest to commence observations of the GW error region and searches in a bluer band. Follow-up in blue filters was particularly advantageous for GW170817, whereby we caught the blue emission from AT2017gfo and were able to observe its evolution within the $u$ and UV bands \citep{eva17}. With these advantages combined with improvements made to our observing strategy within the O3 period, under a repeat of the GW170817 scenario, we would have observed AT2017gfo 5 hr before ground-based observatories \citep{tou17}. Another advantage is that {\it Swift} provides simultaneous coverage of all UVOT fields in X-rays with the XRT. Therefore for all UVOT sources of interest, e.g. the EM counterpart to the GW event or other sources of interest, we can immediately provide optical to X-ray ratios, providing an instantaneous snapshot of the multi-band spectral behaviour. This is an advantage not afforded to any other facility performing regular EM follow-up of GW sources. 

We have no strong evidence to identify any of the transients observed by UVOT in O3 as counterparts to the GW events, consistent with the reports by other IR/optical/UV facilities \citep{and19,gol19,lun19,ack20,and20,ant20,cou20,cou20b,gom20,kas20,pat20,tha20,vie20,ana21,ohg21}. The localisations of the triggers are large (ranging from tens to thousands of square degrees). The observing strategy of {\it Swift} means that a maximum of 800 fields will be observed by UVOT for any given GW event, covering $\sim68\,{\rm deg}^{2}$. In addition, the distances of the BNS merger events are typically a factor of 5 greater than that of GW170817 \citep[$\sim 40$ Mpc; ][]{cou17}, with BH-NS and BBH mergers typically at distances a factor of 10 greater than GW170817. For AT2017gfo, the EM counterpart of GW170817, the UVOT $u$-band magnitude was $18.19^{+0.09}_{-0.08}$ mag upon first detection, 15.3 hr after the trigger and $19.00^{+0.17}_{-0.15}$ mag one day after the trigger. As the typical distance of BNS GW events in the O3 period is a factor of 5 times more distant, this would imply the typical brightness in the $u$-band would be 3.3 magnitudes fainter than was observed for GW170817 \citep[as discussed by][]{ant20}. Therefore, it is not surprising that UVOT did not detect an EM counterpart during O3 since the larger distances and localisations of the GW alerts make any potential EM counterpart harder to detect because a larger portion of the sky has to be searched, and any potential EM counterpart will be much fainter. However, it is worth noting that if an EM counterpart is observed earlier in its evolution, it may be brighter. In O3 {\it Swift} typically began searches for EM counterparts within 2-4 hours after the trigger, 4-8 times earlier than the first UVOT exposure of AT2017gfo. 

\subsection{Serendipitous Sources}
During the follow-up by {\it Swift} of 18 GW events in the O3 period, 27 sources were found by UVOT that are classified as sources of interest. These sources were determined to have changed in magnitude at 3$\sigma$ confidence compared with archival $u$ or $g$-band catalogued values. The date the catalogued entry was observed and the choice of catalogue is dependent on the source location on the sky and when these catalogues were produced. Therefore these sources have been selected heterogeneously. During the O3 period, {\it Swift}/UVOT also followed up a further 13 sources reported by other facilities, which we also include in this discussion. Using catalogue information reported by VizieR, we divided these 40 sources into five initial classifications resulting in 11 candidate AGN/QSOs, 3 CVs, 9 SNe, 11 unidentified sources and 6 uncatalogued sources. We will now explore these sources to try to confirm these classifications and to determine the progenitors of the unidentified and uncatalogued sources. 

Of the candidate AGN, 8 have some form of distance measurement (5 photometric redshift, $\sim0.05-3.7$; 3 spectroscopic redshift, 0.7-1.5), which are typical of AGN distances \citep[from $z\sim 0.01$ to $z > 7$;][]{zhe04,maz17}. Of the nine candidate AGN with archival SDSS observations, eight are classified morphologically as stars through the SDSS photometry suggesting the photometry is dominated (at least at that epoch) by the AGN rather than the galaxy. Their light curves are the best sampled of all our sources. The light curves span several years and have low-level changes at small time scales, and some have larger changes over several years (see Supplementary Figs. S.1-S.3). The low-level changes are consistent with that expected for AGN, which is $<1$ mag and typically between 0 to 0.5 mag \citep{mac12}. This behaviour is consistent with that expected for AGN \citep{ulr97}, whereby the optical AGN temporal variability is often described as damped random walk \citep{kel09}. A few of the candidate AGN in our sample display larger changes in magnitude $>1$ mag over a decade or more in time. Similar behaviour was also identified by \cite{law16} for a small number of blue transients. These transients showed extreme changes in brightness, with changes of $>1.5$ mag occurring over a decade in time. These were spectroscopically confirmed as AGN. \cite{law16} hypothesised that these large-amplitude changes in brightness might be due to changes in accretion state or large amplitude microlensing by stars in foreground galaxies. They note that AGN with these extreme changes in brightness are rare of the order of 1 per 1000 or 10000. This suggests that the candidate AGN in our sample with changes of $>1$ mag over a decade or more in time may also be `hypervariable' AGN. Tidal disruption events (TDEs) may also cause a temporary increase in brightness over several months. With the limited information available and without spectroscopic identification, we can not exclude the possibility that the increase in brightness of one or more of these candidate AGN may actually be due to a TDE. However, the likelihood of a TDE being observed in the UVOT field of view during O3 is minimal, with $<<0.1$ event expected within the entire UVOT O3 coverage (see \S \ref{density}). 

\cite{gra20} suggested ZTF19abanrhr as a possible EM counterpart to GW190521 \citep[c.f.][]{ash20,pal21}. This transient was associated with a flare of an AGN peaking around 50 days after the GW trigger. The light curve of this event before the GW event varied by only a few per cent, while the flare was a $5\sigma$ deviation from the baseline ZTF flux lasting around 50 days. There are 4 UVOT sources of interest, classified as candidate AGN, found during the follow-up of a GW event associated with a BBH merger: Q0\_src10, Q0\_src28, Q0\_src36 (S190412m) and Q1\_src40 (S200224ca). Of these sources, only Q1\_src40 has a distance that could be consistent with that of the GW event. The light curve of this event, shown in Supplementary Fig. S.3, particularly around the time of the GW trigger, is not markedly different to the other candidate AGN displayed in the other panels. However, a flare associated with the BBH merger is likely to present only a few days to weeks after the GW \citep{mck19}. Indeed ZTF19abanrhr was first detected 34 days after the GW trigger \citep{gra20}. Since UVOT observations cease within a few days of the GW trigger, the brightening of an AGN associated with a GW event would likely be missed.

The 11 unidentified sources have light curves spanning similar duration as observed for the candidate AGN, although for most sources, there are fewer observations. These sources fall into three main clusters in Fig. \ref{peak_delta}. The first group includes Q1\_src82 (S190718y), Q1\_src49, Q1\_src54 and Q1\_src113 (S190814bv), Q1\_src56 from S200115j. These are slightly fainter but have a similar magnitude change as the candidate AGN in Fig. \ref{peak_delta}. On the $g-r$ vs $r-i$ colour panel of Fig \ref{Optical_colours}, Q1\_src49, Q1\_src54, Q1\_src113 (S190814bv) are consistent with the colours of the candidate AGN. The similarities in colour, brightness and $\Delta m$ suggests that these sources are also likely AGN. However, Q1\_src82 is inconsistent with the candidate AGN colours. For Q1\_src82, we note that in $g-r$ versus $r-i$ panel of Fig. \ref{Optical_colours}, the archival magnitudes of this source are unusually blue compared to the rest of the sources and highlighted regions. This suggests this source is not an AGN, but with the limited information available, we cannot constrain the origin of this source any further. 

The second cluster of unidentified sources includes Q0\_src186 (S190425z), Q0\_src93 (S190930t) and Q0\_src47 (S200224ca). They are all brighter and have a larger $\Delta m$ compared to the candidate AGN and are positioned close to AT2017gfo, the EM counterpart to GW170817 in Fig. \ref{peak_delta}. Q0\_src186 and Q0\_src47 are similar in their properties. For Q0\_src186, from UVOT observations, we measured a $\Delta m_u >2.0$, but the source is not detected in stacked $g$-band Pan-STARRS exposures with limiting magnitude of this survey typically 23.3 mag. This suggests that $\Delta m$ maybe $>4.5$ magnitudes. An estimate of the rate at which the source faded can be derived from the change in $u$-band magnitude, $\Delta m_u >1.4$, between detection and the next UVOT exposure taken 27 hours later. This gives a lower limit to the decay rate of $1.2$ mag per day. For Q0\_src47, from UVOT observations we measured $\Delta m_u >2.2$. For this source, there are also $g$-band stacked images available in Pan-STARRS. The difference between this and the $u$-band initial detection suggests the source changed in brightness by $\sim5$ mag. We can again estimate the rate of decay of this source by looking at the time and change in brightness between the UVOT detection and the next exposure. The change is $\Delta m_u >1.4$ within 2.5 days. This implies a decline rate of $>0.6$ mag per day. The large decline rates of both Q0\_src186 and Q0\_src47, enables us to rule them out as slow evolving transients such as such as SNe \citep{whe90,gal12}, TDEs \citep{van20} and AGN \citep{mac12,smi18} but we cannot exclude faster evolving transients such as GRBs \citep{sar98}, KNe \citep{met19}, fast-evolving CVs and novae \citep[e.g][]{hac18}, and flare stars \citep{ost10,sch14}. Archival observations of Q0\_src47 and Q0\_src186 by Pan-STARRS show point-like sources suggesting the archival objects are stars or distant galaxies. For both sources, the photometry increases in brightness towards the red. Q0\_src186 is not detected in $g$. Q0\_src47 is at least two magnitudes fainter in $g$ and $r$ compared to $i$, $z$ and $Y$. In the $g-r$ vs $r-i$ colour panel of Fig. \ref{Optical_colours}, the archival photometry of both Q0\_src47 and Q0\_src186 indicates these are the reddest sources in $r-i$. A hypothesis of Q0\_src186 being a flare of a red-brown dwarf was put forward by \cite{GCN:24326}. Q0\_src93 is more peculiar. To date, it continues to be detected by UVOT but is decreasing in brightness. Archival observations by DES place this source as the reddest object on the $g-r$ vs $r-i$ colour plane, redder than galaxies and stars. A more thorough investigation of this source will be presented in Oates et al., (in prep). 

The third cluster of unidentified sources are Q1\_src1, Q1\_src39 and Q1\_src62, all from S200115j. These sources lie in the top left of Fig. \ref{peak_delta}, they are faint but have $\Delta m_u >4$. For Q1\_src1, it is difficult to put meaningful constraints on the potential rate of decay of this source from UVOT observations since only two short exposures with limits consistent with the detection magnitude were taken after the initial UVOT detection, both within the first 24hr. UVOT took a third longer exposure, but only five months later. This source was not detected in a short exposure taken 1.5 hr before the initial detection, though the limit is only slightly deeper than the detection; this could suggest a rapid onset. However, in archival Pan-STARRS photometry, Q1\_src1 shows a rapid change in $i$-band of 2 magnitudes on a timescale of 20 min. For Q1\_src39, we cannot give a very constraining decay rate either since there are ten days between detection by UVOT and the subsequent exposure, but this event must have had a reasonably rapid onset since the change from an image taken 20 hours prior is $\Delta m_u >0.9$. For Q1\_src62, it is also not possible to provide strong constraints on the onset or decay rate. The closest UVOT exposures to the UVOT detection exposure were taken 16 hours before and ten days after and have $3\sigma$ upper limits consistent with the detection magnitude. In the $g-r$ vs $r-i$ panel, we only have a lower limit to the redness of Q1\_src1 since we only have an upper limit in $r$ for this source, but it is one of the reddest sources in this panel. For Q1\_src39 and Q1\_src62, these two sources are among the reddest sources in $g-r$, though the errors are large. These sources are very faint in archival images, in Pan-STARRS for Q1\_src1 and CFHTLS for Q1\_src39 and Q1\_src62, suggesting a faint star or distant galaxy origin. With the red archival values and the potentially rapid onset of Q1\_src1 and Q1\_src39, we suggest Q1\_src1, Q1\_src39, Q1\_src62 could have similar nature as Q0\_src47 (S200224ca) and Q0\_src186 (S190425z).

Six of the sources of interest we initially classified as uncatalogued. For these sources, there is no archival information providing an identification, and for only one, we were able to produce a light curve: ZTF20aamvmzj, shown in Supplementary Fig. S.7. In this figure, we display data obtained by both UVOT and \cite{kas20}. Blue evolution is observed, with the UV filters declining while the ZTF $g$ and $r$-band remain constant. Power-law fits also confirm this, which give $\alpha = 0.04$ in the $r$-band \cite{kas20} and $-1.45\pm0.23$ measured from the combined UVOT data. ZTF20aamvmzj displays behaviour reminiscent of Type II SNe. In Supplementary Fig. S.7, we overlay the optical/UV light curves of ASASSN-14ha \citep{val16}, which were obtained from the Open Supernova Catalog \citep{gui17}. We scaled the light curves of ASASSN-14ha by six magnitudes, corresponding to a factor $\approx 16$ in distance. The similarities in light curve evolution between ZTF20aamvmzj and ASASSN-14ha suggests ZTF20aamvmzj is most likely a Type II SN at a distance $\sim350 $Mpc.

For Cand-A09 we can not draw any conclusion as to the rate of decay of this object since the second UVOT image was taken 2 years after the initial detection. However, we note that in Fig. \ref{peak_delta}, this source lies just beneath the CV cluster and only a lower limit is known on $\Delta m$. Therefore it is possible that this source is a CV.

For the remaining four uncatalogued sources, Q0\_src136 (S190425z), Q1\_src33 (S190930t), Q1\_src58 (S200115j) and Q1\_src54 (S200224ca), we re-examined the UVOT images to ensure these sources were not due to scattered light. If these were ghosts, we would expect a bright source in the FOV. However, no bright sources are apparent in any of the images. The full-frame image and zoom-in on the source location for each of these objects are provided in Supplementary Fig. S.8. In Fig. \ref{peak_delta}, three of these sources have peak magnitudes slightly brighter than that found for most of the candidate AGN and the unidentified sources, and $\Delta m$ for these three sources is $\gtrsim1$mag. The first image taken after the initial UVOT detection for two of these sources, Q0\_src136 (S190425z) and Q1\_src54 (S200224ca), was over two months later, so we cannot draw useful constraints on the decay rate or hence draw further conclusions on the nature of these two sources. Q1\_src54 does, however, have a WISE source within $4\arcsec$, which may indicate an underlying red object. Interestingly, for Q1\_src58, we have a detection at a magnitude of $21.31\pm0.37$ lasting 486s and then another 476 exposure starting just 48s after the first exposure giving a $3\sigma$ upper limit ($>21.22$). However, for this source and Q1\_src33 (S190930t), the detection and upper limit measured in the subsequent exposure are consistent, so we can not draw firm conclusions. We suggest that these four uncatalogued UVOT sources could be similar to Q0\_src47 (S200224ca) and Q0\_src186 (S190425z). We base this assumption on the lack of archival detections and because these sources were detected in a single UVOT exposure, similar to Q1\_src1, Q1\_src39 and Q1\_src62 (S200115j).

\subsubsection{Transient Density in UVOT}
\label{density}
Within the 424 deg$^2$ observed at least once by UVOT during the O3 period, UVOT found 27 transient sources. These were deemed to be transient based on their increase in brightness at 3$\sigma$ confidence compared with archival $u$ or $g$-band catalogued magnitudes. This provides a detection surface density of transient sources brighter than 21.2 mag of 0.064 deg$^{-2}$ in the $u$ filter. Of the 27 sources, we initially identified 10 as candidate AGN, and 5 unidentified sources are also likely to be AGN, bringing the total of candidate AGN to 15. The rate of AGN with transient behaviour found by the UVOT during the O3 period is therefore 0.024 - 0.035 deg$^{-2}$ in the $u$ filter. Similarly, we estimate that between 5-9 of the UVOT sources of interest are fast evolving transients, which are red sources in quiescence, e.g. Q0\_src47 (S200224ca) and Q0\_src186 (S190425z), one possibility is that these are flare stars. The transient density for these sources is 0.012 - 0.021 deg$^{-2}$ in the $u$ filter. We did not conclusively detect any TDE candidates. The TDE rate brighter than $\sim21$ mag is 1 per month \citep{van20b}, assuming a duration of six months, there should be a few TDEs on the sky at any time. With the UVOT sky coverage of 424 deg$^2$ during O3, we, therefore, expect a rate of a few $\times0.01$ TDEs within the O3 coverage. This is consistent with zero confirmed TDEs observed by UVOT.

\subsection{Future Improvements to GW Follow-Up by UVOT}
{\it Swift}/UVOT was essential to the discovery of blue emission produced by the KN associated with 170817 \citep{eva17}. 
Through the rapid planning capabilities devised by the {\it Swift} team \citep{tou17}, the typical start time of {\it Swift} tiling is 2-4 hr after the trigger, with observations commencing as early as 1.8hr for S200114f. This rapid capability combined with the relatively unrestricted observing windows and the blue coverage of the UVOT is important for detecting the KN and observing its early evolution \citep{met15,met19}. With the advantages {\it Swift} has over ground-based facilities, the {\it Swift} team are continually endeavouring to optimise the search of the LVC error regions. 

One of the greatest drawbacks UVOT currently has is the identification of new sources within galaxies. The majority of thumbnails produced by the UVOT pipeline are marked as `gal'. Currently, these galaxy images are manually scrutinised for changes in brightness or any new point sources by comparing with the archival UVOT image, if available, or the DSS image. However, the capacity to find transients this way is limited. We do not have archival UVOT images for most of these galaxies. Therefore a visual inspection and comparison to DSS images in the blue band can easily miss small changes in brightness and perhaps even large changes if the galaxy core is overexposed in the DSS. One of the improvements that attempts to address this will be the completion of the SGWGS catalogue (as discussed in \S \ref{SGWGS}; Tohuvavohu et al., in prep), which is expected to observe a total of 13000 galaxies. Having a template UVOT, $u$-band image to directly compare against a GW $u$-band tile should mitigate some of the issues of comparing against archival images from other facilities. These $u$-band images and other archival UVOT $u$-band images will be used during side-by-side (or blinked) comparison with the tiled GW images, and we shall explore whether image subtraction improves the identification of the transient source and, if successful, implement this in the UVOT pipeline.

With experience gained during O3, we are currently looking at how we can improve the UVOT pipeline and manual checking procedures in the run-up to O4 to optimise the possibility of finding an EM counterpart in the optical/UV. We will expand the UVOT pipeline to automatically create light curves for all Q0, Q1, Q2 and Q3 sources, using UV/optical/IR catalogues and photometry available through VizieR, IRSA and other sources. The pipeline will automatically report the listing of sources in astronomical catalogues as potential AGN, RR lyre stars etc. Using archival information, we will be able to quickly check the position of the source on colour-colour diagrams such as the $g-r$ vs $r-i$ in Fig. \ref{Optical_colours}, in order to prioritise follow-up for those UVOT transients. In particular, UV transients consistent with residing in elliptical galaxies may be worth prioritising since, according to Fig. \ref{Optical_colours}, UV transients appear to be relatively rare in elliptical galaxies. There is some chance that short GRBs may be hosted in elliptical galaxies. All of this will enable rapid identification of sources, identify whether they are known objects and will initiate rapid follow-up to determine if the source is a candidate EM counterpart of the GW event or another source of interest that should be explored further. 

In addition, for around 10 per cent of tiles observed by {\it Swift}, UVOT uses a less sensitive filter (e.g. $uvw1$) or a blocked filter so that damage to the UVOT is avoided due to bright stars/fields. This means that for every GW event observed by {\it Swift}/UVOT, there are a non-negligible number of potential host galaxies that can not be observed. In the run-up to O4, we will explore whether we can adapt the field selection algorithm to offset the pointing, where possible, such that the bright star is avoided, but the potential host galaxy within a given tile is still observed.

\section{Conclusion}
\label{conclusions}
In this paper, we summarised the {\it Swift}/UVOT pipeline and follow-up of Gravitational Wave events in the third LIGO-Virgo observing run, O3. During this cycle {\it Swift}/UVOT GW followed up 18 GW events to differing extents. Using the classifications issued by LVC and reported in the GraceDB, these events can be categorised as 5 BBH mergers, 6 BNS mergers (three of which had a larger probability of being terrestrial rather than BNS), 3 BH-NS, two Mass Gap triggers and two unmodelled/burst triggers. Of these, three were quickly retracted: 1 BNS, 1BH-NS and 1 unmodelled and three were not included in the GWTC-2 catalog of O3a: 2 BNS and 1 BH-NS. Across all 18 GW events, {\it Swift}/UVOT observed 6441 individual fields (tiles). All images were processed through the UVOT GW pipeline, which identifies sources of interest and gives them a quality flag. All sources identified are manually verified. Sources deemed more likely to be actual transients are assigned a flag of 0 (referred to as Q0 sources) or 1 depending on their magnitude. Sources dimmer than a magnitude of 19.9 (a conservative sensitivity limit to obtain a signal to noise $>$5 in the $\sim 80s$ tiling observations) are assigned a value of 1. The automatic analysis pipeline found a total of 1516 Q0 and Q1 sources to be verified.

After manual inspection, 27 sources were considered to be of interest. These sources were determined to have changed in magnitude at 3$\sigma$ confidence compared with archival $u$ or $g$-band catalogued values. {\it Swift}/UVOT also followed up a further 13 sources reported by other facilities during the O3 period. Using catalogue information reported by VizieR, we divided these 40 sources into five initial classifications: 11 candidate AGN/QSOs; 3 Cataclysmic Variables; 9 supernovae; 11 unidentified sources with archival photometry, but no classification and 6 uncatalogued sources for which no archival photometry was available. We further examined these sources of interest to determine if we could classify the unidentified and uncatalogued sources. For this analysis, we examined and compared the information available from all 40 sources. We find it likely that most of the unidentified and uncatalogued sources are AGN, a class of fast-evolving transient and one source may be a CV. We have no substantial evidence to identify any of these transients as counterparts to the GW events.

Finally, we suggested improvements that can be made to the UVOT pipeline and manual verification of sources before the start of the LIGO-Virgo-KAGRA O4 run. One of the most significant improvements will be incorporating archival UVOT images into the pipeline for comparison with UVOT images taken during the O4 run, likely commencing mid-2022. In particular, we will be including 13000 images from the SGWGS catalogue, a catalogue of $u$-band images of nearby galaxies. UVOT is a crucial instrument in the detection and follow-up of blue emission associated with GW events. {\it Swift}'s unique capabilities place it in a prime position to detect the EM counterpart to a GW early on and to provide the astronomical community with coverage of large portions of the GW error region that may otherwise be constrained or have delayed observations from ground-based telescopes. The current estimated reentry time of
{\it Swift} is around 2035, providing funding is continued, this suggests UVOT has many more successful years ahead chasing EM counterparts to GWs as well as remaining the workhorse of the transient community.

\section{Acknowledgments}
We thank the referee and M. Nicholl for useful discussion and suggestions that improved the paper. This research has made use of data obtained from the High Energy Astrophysics Science Archive Research Center (HEASARC) and the Leicester Database and Archive Service (LEDAS), provided by NASA's Goddard Space Flight Center and the School of Physics and Astronomy, University of Leicester, UK, respectively. This research has also made use of a number of public services: the VizieR catalogue access tool, CDS, Strasbourg, France (DOI: 10.26093/cds/vizier); the International Astronomical Union Minor Planet Center, which is hosted by the Center for Astrophysics at the Harvard \& Smithsonian and is funded by NASA; data from the European Space Agency (ESA) mission {\it Gaia} (\url{https://www.cosmos.esa.int/gaia}), processed by the {\it Gaia} Data Processing and Analysis Consortium (DPAC, \url{https://www.cosmos.esa.int/web/gaia/dpac/consortium}). 
The CSS survey is funded by the National Aeronautics and Space Administration under Grant No. NNG05GF22G issued through the Science Mission Directorate Near-Earth Objects Observations Program. The CRTS survey is supported by the U.S.~National Science Foundation under grants AST-0909182 and AST-1313422. This publication makes use of data products from the Wide-field Infrared Survey Explorer, which is a joint project of the University of California, Los Angeles, and the Jet Propulsion Laboratory/California Institute of Technology, funded by the National Aeronautics and Space Administration. The CFHTLS data is based on observations obtained with MegaPrime/MegaCam, a joint project of CFHT and CEA/IRFU, at the Canada-France-Hawaii Telescope (CFHT) which is operated by the National Research Council (NRC) of Canada, the Institut National des Science de l'Univers of the Centre National de la Recherche Scientifique (CNRS) of France, and the University of Hawaii. This work is based in part on data products produced at Terapix available at the Canadian Astronomy Data Centre as part of the Canada-France-Hawaii Telescope Legacy Survey, a collaborative project of NRC and CNRS. This research has made use of the NASA/IPAC Infrared Science Archive, which is funded by the National Aeronautics and Space Administration and operated by the California Institute of Technology. The Pan-STARRS1 Surveys (PS1) and the PS1 public science archive have been made possible through contributions by the Institute for Astronomy, the University of Hawaii, the Pan-STARRS Project Office, the Max-Planck Society and its participating institutes, the Max Planck Institute for Astronomy, Heidelberg and the Max Planck Institute for Extraterrestrial Physics, Garching, The Johns Hopkins University, Durham University, the University of Edinburgh, the Queen's University Belfast, the Harvard-Smithsonian Center for Astrophysics, the Las Cumbres Observatory Global Telescope Network Incorporated, the National Central University of Taiwan, the Space Telescope Science Institute, the National Aeronautics and Space Administration under Grant No. NNX08AR22G issued through the Planetary Science Division of the NASA Science Mission Directorate, the National Science Foundation Grant No. AST-1238877, the University of Maryland, Eotvos Lorand University (ELTE), the Los Alamos National Laboratory, and the Gordon and Betty Moore Foundation. SRO gratefully acknowledges the support of the Leverhulme Trust Early Career Fellowship. AAB, NPMK, MJP, KLP, PAE, APB and JPO acknowledge funding from the UK Space Agency. MDP acknowledges support for this work by the Scientific and Technological Research Council of Turkey (T\"UBITAK), Grant No: MFAG-119F073. EA, MGB, SC, GC, AD, PDA, AM and GT acknowledge funding from the Italian Space Agency, contract ASI/INAF n. I/004/11/4. This work is also partially supported by a grant from the Italian Ministry of Foreign Affairs and International Cooperation Nr. MAE0065741. PDA acknowledges support from PRIN-MIUR 2017 (grant 20179ZF5KS). DBM is supported by research grant 19054 from Villum Fonden.

\section{Data Availability}
The data underlying this article are available in the {\it Swift} archives at https://www.swift.ac.uk/swift\_live/, https://heasarc.gsfc.nasa.gov/cgi-bin/W3Browse/swift.pl, https://www.ssdc.asi.it/mmia/index.php?mission=swiftmastr, and in the online supplementary material.


\bibliographystyle{mn2e}   
\bibliography{Swift_UVOT_GW_O3} 

\IfFileExists{\jobname.bbl}{}
 {\typeout{}
  \typeout{******************************************}
  \typeout{** Please run "bibtex \jobname" to optain}
  \typeout{** the bibliography and then re-run LaTeX}
  \typeout{** twice to fix the references!}
  \typeout{******************************************}
  \typeout{}
 }

\end{document}